\documentclass[journal]{IEEEtran}
\usepackage{upgreek}
\usepackage{graphicx}
\usepackage{subfigure}
\usepackage{amsmath,bm}
\usepackage{amssymb} 
\usepackage{mathrsfs} 

\usepackage{enumerate}

\usepackage[square, comma, sort&compress, numbers]{natbib}

\usepackage{color}
\newcommand{\mll}{\color{blue}}
\newcommand{\mrr}{\color{black}}

\ifCLASSINFOpdf
\else
\fi

\begin{document}
	
	\title{\huge Time-Domain Analysis for Resonant Beam Charging and Communications With Delay-Divide Demodulation}
	\author{Mingliang Xiong, Shun Han, Qingwen Liu,~\IEEEmembership{Senior Member,~IEEE}, and Shengli Zhou,~\IEEEmembership{Fellow,~IEEE}
		
		\thanks{
			M. Xiong, Q. Liu, and S. Han are with the College of Electronics and Information Engineering, Tongji University, Shanghai 201804, China (e-mail: xiongml@tongji.edu.cn;    qliu@tongji.edu.cn; hanshun@tongji.edu.cn).}
		\thanks{S. Zhou is with the Department of Electrical and Computer Engineering, University of Connecticut, Storrs, CT 06250, USA (e-mail: shengli.zhou@uconn.edu).}
	}

	\maketitle
	
	\begin{abstract}
		Laser has unique advantages such as abundant spectrum resources and low propagation divergence in wireless charging and wireless communications, compared with radio frequency. Resonant beams, as a kind of intra-cavity laser beams, have been proposed as the carrier of wireless charging and communication, as it has unique features including high power, intrinsic safety, and self-aligned mobility. \mll{}However, this system has problems such as  intra-cavity echo interference and power fluctuation.\mrr{}
		To study the time-domain behavior of the resonant beam system, we create a simulation algorithm by discretizing the laser rate equations which model the dynamics of the excited atom density in the gain medium and the  photon density in the cavity. The simulation results are in good agreement with  theoretical calculation. We also propose a delay-divide demodulation method to address the echo interference issue, and use the simulation algorithm to verify its feasibility. The results  show that the resonant beam charging and communication system with the proposed demodulator is feasible and performs well. The analysis in this work also helps researchers to deeply understand the behavior of the resonant beam system.
	\end{abstract}
	
	

\section{Introduction}\label{sec:intro}
\IEEEPARstart{I}{nformation} and power are two significant requirements of most mobile electronic devices. These requirements are becoming more and more urgent with the continuing development of new technologies and  living standard. For instance, online meeting/class rooms in the future are expected to use virtual reality~(VR) or augmented reality~(AR) technologies to move the attenders to a metaverse where they can experience face-to-face chat even if they are apart from thousands of kilometers~\cite{a211015.01, a211015.03, a211015.02}. However, the three-dimensional~(3D) holographic images/videos transmitted between VR/AR devices have much greater data size than those in two-dimensional (2D) scenario. Moreover, to transmit and process large amount of data, the energy consumption increases dramatically as well.
Thus, in the future, the six-th generation (6G) or beyond 6G mobile network is expected to support both information and power transfer~\cite{a180820.09}.

Resonant beam charging and communications systems have been proposed to support simultaneously transferring power and information to  remote mobile devices~\cite{a180727.01,a200528.01,a200508.02}. This technology is based on a spatially separated laser resonator~(SSLR). Both the transmitter and the receiver are the components of the SSLR, different from an integrated laser transmitter. Although integrated laser communication systems have many advantages in spectrum resource, bandwidth, and transmission distance, they still face challenges in some specific scenarios such as indoor mobile communications. Safety and mobility are two important issues in this scenario. As for safety, making a low-power safeguard/detection beam around the high-power laser is one of the protection methods for laser power transfer~\cite{a180805.04}, while it  needs accurate positioning/tracking system and sensitive detection system. In terms of mobility, there are many outstanding technologies for fast directing narrow laser beams to remote devices, such as
fiber arrays~\cite{a191111.01}, crossed diffraction gratings~\cite{a190611.03}, optical phased arrays~(OPAs)~\cite{a201201.02}, and spatial light modulators~(SLMs)~\cite{a190514.01}, but the alignment accuracy and the tracking speed also depend on the performance of the receiver positioning technologies~\cite{a210901.06}. \mll{}In addition to wireless communications, laser-based wireless power transfer~(WPT) also faces  challenges in receiver positioning and tracking.\mrr{}

\begin{figure}
	\centering
	\includegraphics[width=3.4in]{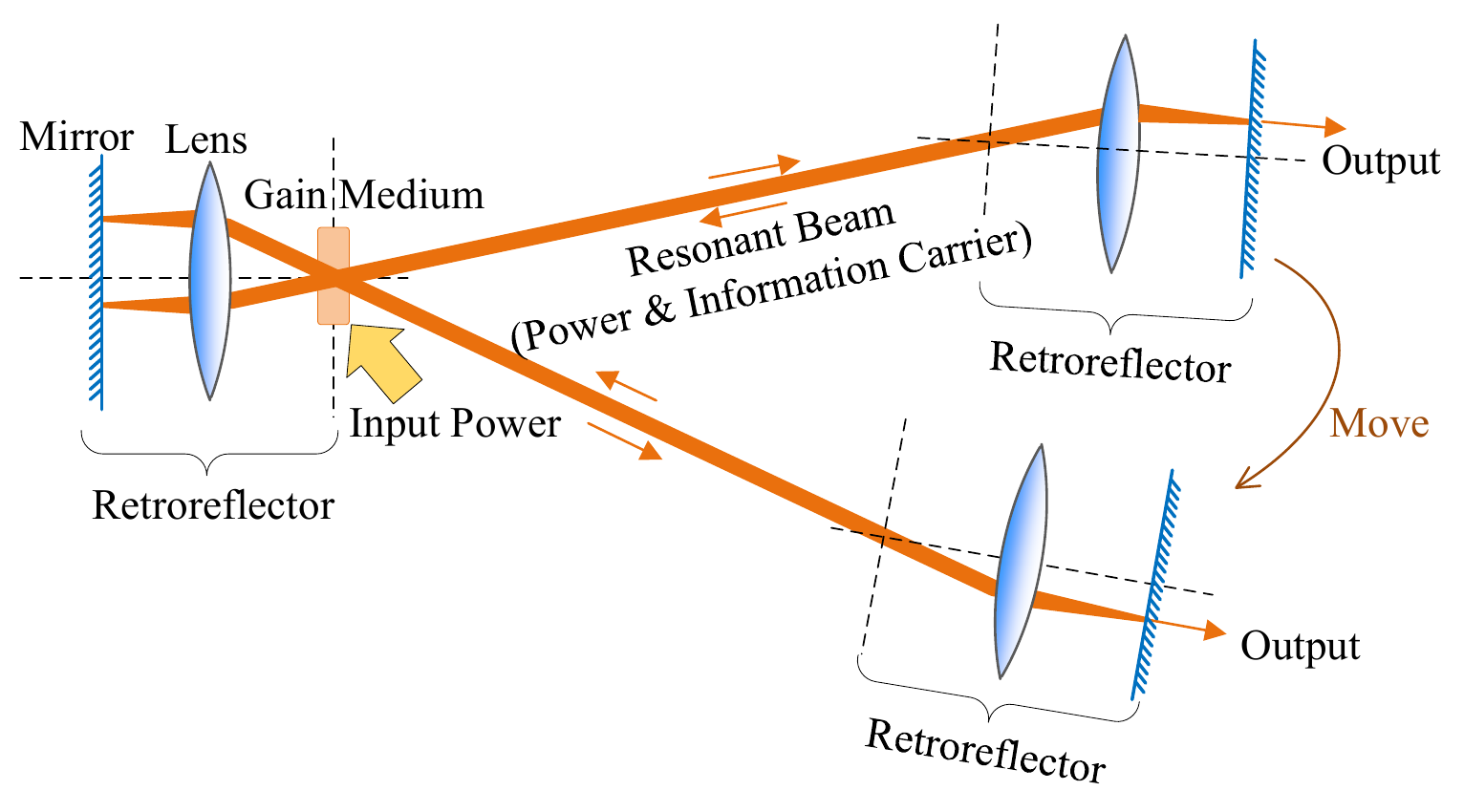}
	\caption{Mechanism of resonant beam generation}
	\label{fig:mechanism}
\end{figure}

Resonant beam charging and communications based on SSLR can concurrently support intrinsic safety and self-aligned mobility. The primary structure of SSLR originates from Linford's very long laser proposed in 1973, which consists of two corner cubes and a gain medium~\cite{a190318.02, a190318.01}. Corner cubes are retroreflectors which reflect beams to their sources. Other retroreflectors such as cat's eye and telecentric cat's eye can also be used in the SSLR. As shown in Fig.~\ref{fig:mechanism}, an SSLR consisting of two telecentric cat's eye retroreflectors can force photons to oscillate in its cavity even if the position and pose of the retroreflectors are changed; this process generates a light beam connecting the two ends. The gain medium amplifies the beam power, compensating for the power loss that the beam experienced during oscillation. The company Wi-charge discloses a wireless charging system design based on SSLR (also known as distributed laser cavity), in which two telecentric cat's eye retroreflectors~(TCRs) are employed to replace the corner cubes~\cite{wicharge}. \mll{}A focal TCR~(FTCR) was proposed in~\cite{a211015.06}; in this paper,\mrr{} the stable regime of the FTCR-based SSLR was found by theoretical derivation, which is an improvement as the stable regime provides very low diffraction loss within a long moving range. \mll{}The experiment design of SSLR-based mobile charging   in~\cite{liu2021mobile} demonstrated\mrr{} charging a smartphone within $2$-m distance and $6^\circ$ field-of-view~(FOV) with more than $0.6\mbox{-W}$ electrical power (received more than $5$-W optical power). An experiment on adjustable-free and movable TCR-based SSLR was also demonstrated in~\cite{a210831.01}. Besides, the safety and mobility of the SSLR were verified by spatial wave-optical simulation~\cite{a211015.04,a210823.02}. An enhanced experiment on the TCR-based SSLR achieved $5$-m transmission distance and $5.91\mbox{-W}$ received optical power. Moreover, a new SSLR structure based on spatial wavelength division was proposed in~\cite{a190926.02}, which realized $1.7$-mW received power at $1$-m distance. Besides, capacity analysis of kilometer-level  mobile resonant beam communications is presented in~\cite{a210901.01}.

However, there are two significant issues in the resonant beam charging/communication system: the echo interference in modulation and the power fluctuation in movement~\cite{a211015.07}. These issues affect the efficiency of battery charging and the feasibility of high-capacity communication. A design based on intra-cavity second harmonic generation~(SHG)  was proposed to avoid the echo interference in the resonant beam communication system~\cite{MXiong2021.InSHG}, while the SHG beam lies in the visible spectrum, which may not be accepted in daily life.
Experimentally observing the fast fluctuation of the resonant beam power using an oscilloscope is very expensive, as the frequency band of the echo is much wider than that of the source signal. Time-domain simulation algorithm/program can help reduce the costs of this research, while most related works lack the interest of observing  such a micro timescale which is even shorter than the time for a photon to pass through an optical device~\cite{a210706.01,a181220.01,JZhou2021}. That's why we have never known what the echo interference looks like. Also, there is no existing method for extracting the source information from the chaotic output directly.

The contributions of this work are as follows:
\begin{enumerate}
	\item[\bf 1)] We propose a  simulation algorithm for the resonant beam charging and communication system to analyze the system behavior in the time domain. We  model the dynamics of the excited atoms in the gain medium and the intra-cavity photons by discretizing the rate equations. The simulation algorithm supports observing small-scale power variations, even if the duration is smaller than the time for a photon to pass through the gain medium. We compare the simulation results at the stable state with  theoretical computations to verify the high accuracy of the proposed algorithm.
	\item[\bf 2)] To overcome the echo interference, we propose a delay-divide demodulation method. This method can directly extract the source information from the chaotic received signal without extra optical devices. Using the simulation algorithm, it is easy to examine the feasibility and the performance of the proposed demodulation method. Several important scenarios are simulated, such as the start-up of the pump (driving power), the foreign object intrusion/leaving, and the intra-cavity modulation. We also study the demodulator's performance by counting the bit error rate~(BER) using the on-off keying modulation.
\end{enumerate}

The remainder of this paper is structured as follows.
Section~\ref{sec:model} describes the system model of the resonant beam system, including the static output power model and the dynamic rate equations; it also presents the theoretical foundation of the delay-divide demodulation method.
Section~\ref{sec:imple} demonstrates the implementation of the simulation algorithm as well as the design of modulation and demodulation. In Section~\ref{sec:evalu}, the simulation results are studied, including the responses to the modulation actions and the performance of communication. Finally, conclusions
are drawn in Section~\ref{sec:con}.

\section{System Model}
\label{sec:model}

\begin{figure*}[ht]
	\centering
	\includegraphics[width=5.6in]{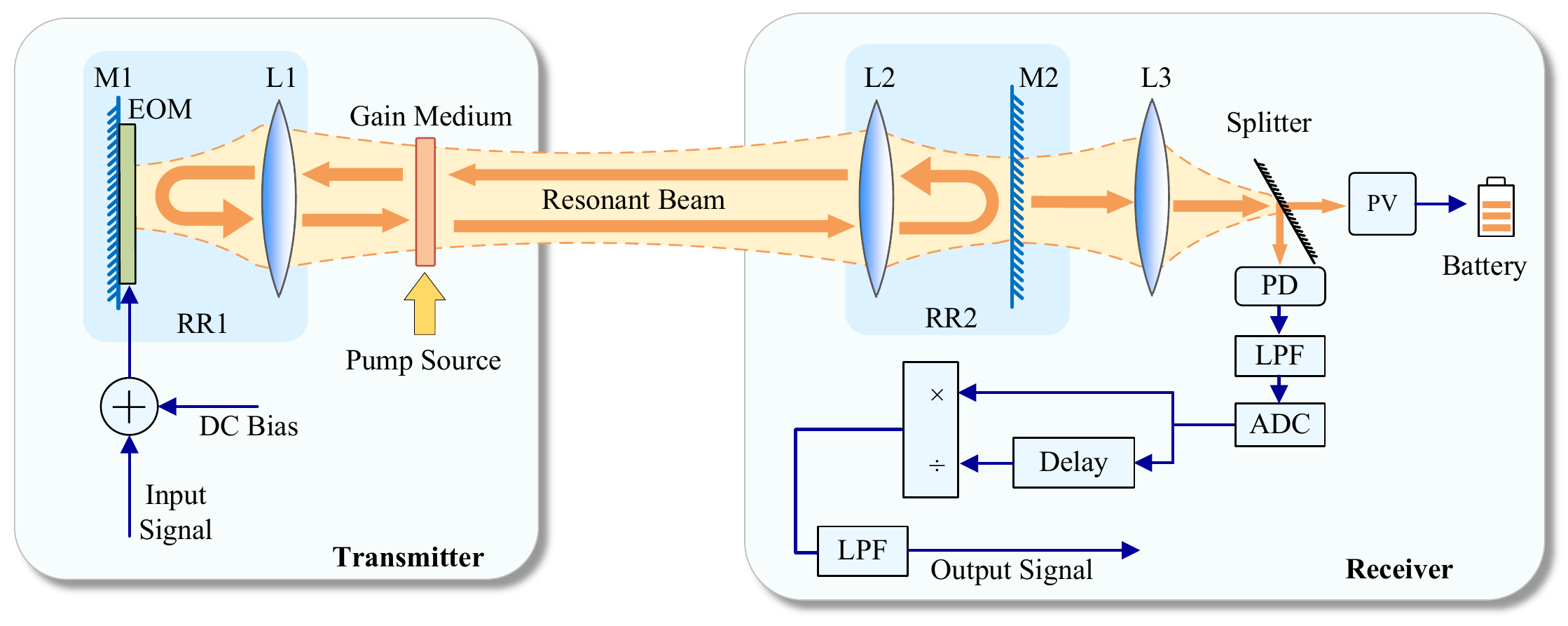}
	\caption{System design of a resonant beam charging and communications system (M1 and M2: high-reflectivity mirrors; L1--L3: Lenses; \mll{}EOM: electro-optic modulator for beam intensity modulation;\mrr{} RR1 and RR2: retroreflectors; PV: photovoltaic panel; PD: photon detector; LPF: lowpass filter; ADC: analog-to-digital converter)}
	\label{fig:design}
\end{figure*}

Our system design is depicted in Fig.~\ref{fig:design}. In this section, we first describe the SSLR system. Then, we present the dynamic model of the intra-cavity resonant beam. Finally, we propose the demodulation method which can overcome the effect of the echo interference.

\subsection{System Description}

Basically, the SSLR consists of two FTCRs RR1 and RR2 -- one at the transmitter and the other at the receiver. Namely, the two ends constitute a resonant cavity. Photons in the cavity oscillate circularly, forming a resonant beam. An external pump source provides the driving power. Generally, the pump source is a diode laser module, if the gain medium is a crystal. There are  other kinds of materials that can be used as the gain medium. For example, semiconductors can be pumped by electricity directly. The original photons in the cavity are generated from the spontaneous emission process of the gain medium. As these photons pass through the gain medium in each round trip, their quantity can be increased by the stimulated emission of the gain medium, and this process compensates for their loss induced by the absorption/diffraction of the optical devices. If the input power is sufficient, the optical amplification and the loss will finally reach a balance where the resonant beam can persist. The resonant beam intensity is modulated by an electro-optic modulator~(EOM) placed between the mirror M1 and the lens L1 to carry information to the receiver. M2 is a partially reflective mirror, so a portion of the photons can pass through M2 and then be collected for battery charging and information demodulation. The splitter determines how much power to be allocated for  charging, and the remainder is for demodulation. The demodulation process contains two low-pass filters~(LPFs), an analog-to-digital converter~(ADC), a delay unit, and a divider. More details about the demodulation is presented in the next section.

\begin{figure}
	\centering
	\includegraphics[width=2.9in]{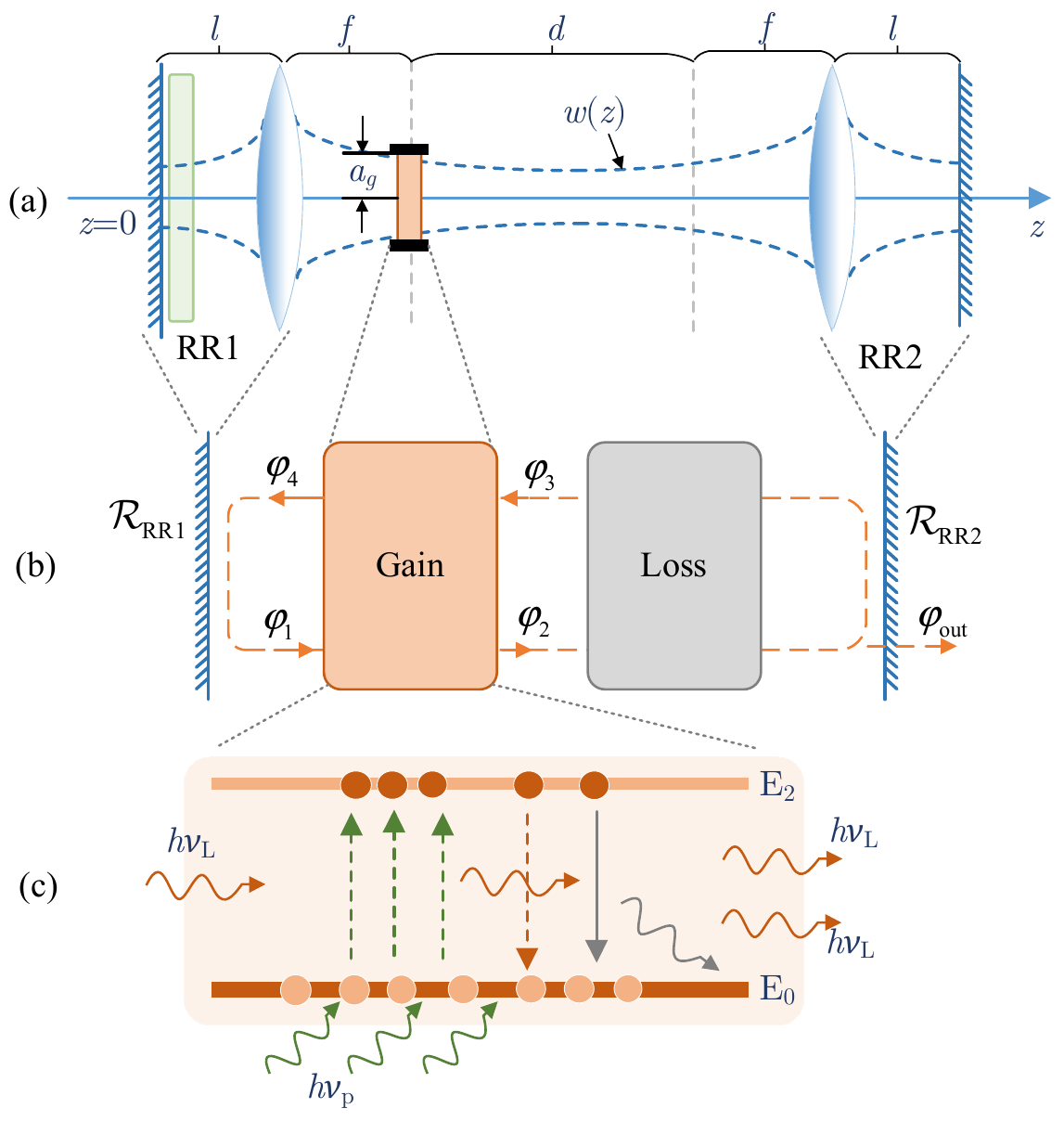}
	\caption{Diagram of the cavity and the gain medium: (a) cavity structure and beam radius along the optical axis $z$; (b) photon circulation in the cavity; and (c) mechanism of optical amplification provided by the gain medium}
	\label{fig:gain}
\end{figure}

Next, we introduce the stable regime and the output power model of the SSLR. A laser resonator should operate in the stable regime where rays will always oscillate in the cavity and can not escape (geometry optics theory). As a result of this characteristic, stable resonator exhibits extremely low diffraction loss. Generally, spherical-mirror resonators with limited cavity lengths are stable. Plane-parallel resonators which consist of two parallel mirrors are not stable, since only the rays perpendicular to the mirrors can be captured by the resonators.  In geometry optics, an optical element/system can be described by a ray-transfer matrix~\cite{a200522.04}. The matrix of a cascaded system is written by the production of the matrices of the components with the reverse order against the input ray's propagation direction. As demonstrated in Fig.~\ref{fig:gain}(a), for an FTCR, the space interval between the mirror and the lens is $l$. We set the outer focal plane of the lens as the input/output~(IO) plane, namely, the space interval between the lens and the IO plane is equal to the focal length, $f$, of the lens. The space interval between two IO planes is defined as the transmission distance $d$. Then, the single-pass ray-transfer matrix of the FTCR-based SSLR yields~\cite{a211015.06}
\begin{align}
\begin{bmatrix}
A&B\\C&D
\end{bmatrix}
=&	\begin{bmatrix}
1&0\\0&1
\end{bmatrix}
\begin{bmatrix}
1&l\\0&1
\end{bmatrix}
\begin{bmatrix}
1&0\\ -1/f&1
\end{bmatrix}
\begin{bmatrix}
1&2f+d\\0&1
\end{bmatrix}	
\nonumber
\\
&\begin{bmatrix}
1&0\\ -1/f&1
\end{bmatrix}
\begin{bmatrix}
1&l\\0&1
\end{bmatrix}
\begin{bmatrix}
1&0\\0&1
\end{bmatrix}.
\label{equ:ABCD}
\end{align}
By calculation, the elements of the ABCD matrix in~\eqref{equ:ABCD} are obtained as follows:
\begin{equation}
\left\{
\begin{aligned}
A&=-1-\dfrac{d}{f}+\dfrac{dl}{f^2},\\
B&=2f-2l+d-\dfrac{2dl}{f}+\dfrac{dl^2}{f^2},\\
C&=\dfrac{d}{f^2},\\
D&=-1-\dfrac{d}{f}+\dfrac{dl}{f^2}.\\
\end{aligned}
\right.
\end{equation}

Let $g_1^*=A$, $g_2^*=D$, a stable SSLR should satisfy the condition $0<g_1^*g_2^*<1$. For such a symmetric SSLR, the up boundary of the transmission distance can be obtained from  this stable condition; that is $d<4f_{\rm RR}$. The intra-cavity resonant beam consists of two parts -- the leftward-traveling wave and the rightward-traveling wave. Generally, there are multiple transverse modes in the resonant beam if the devices apertures are much bigger than the radius of the fundamental mode TEM$_{00}$. The mode radius of TEM$_{00}$ is obtained as~\cite{a181221.01}
\begin{equation}
w_{00}(z)=\sqrt{-\dfrac{\lambda}{\pi\Im\left[1/q(z)\right]}},
\end{equation}
where $\lambda$ is the wavelength of the resonant beam, $\Im[\cdot]$ takes the imaginary part of a complex number. The $q(z)$-parameter is obtained as~\cite{a211015.06}
\begin{equation}
q(z)=
\left\{
\begin{aligned}
&j |L^*|\sqrt{\dfrac{g_2^*}{g_1^*(1-g_1^*g_2^*)}}+z, ~~~~~~z\in[0,z_{\rm L1}],\\
&\frac{q(z_{\rm L1})}{{-q(z_{\rm L1})}/{f_1}+1}+(z-z_{\rm L1}),~ z\in(z_{\rm L1},z_{\rm L2}],\\
&\frac{q(z_{\rm L2})}{{-q(z_{\rm L2})}/{f_2}+1}+(z-z_{\rm L2}),~ z\in(z_{\rm L2},z_{\rm M2}],\\
\end{aligned}
\right.
\label{equ:qParam}
\end{equation}
where $j=\sqrt{-1}$; $L^*=B$; and $z_{\rm M1}$, $z_{\rm L1}$,  $z_{\rm L2}$, and $z_{\rm M2}$ are the location of M1, L1,  L2, and M2, respectively. The ratio of   the resonant beam radius to the   TEM$_{00}$ mode radius is a constant at any location on the $z$-axis. This ratio is called the beam propagation factor. The beam  radius is determined by the smallest aperture in the SSLR. Generally, the gain medium has the smallest aperture. Hence, the  radius of the resonant beam at location $z$ is obtained as~\cite{a181221.01}
\begin{equation}
w(z)=\dfrac{a_{\rm g}}{w_{00}(z_{\rm g})}\sqrt{-\dfrac{\lambda}{\pi\Im\left[1/q(z)\right]}},
\label{equ:radius}
\end{equation}
where $a_{\rm g}$ is the radius of the gain medium aperture, and $z_{\rm g}$ is the location of the gain medium.

Below are loss factors in the SSLR, and they attenuate the circulating power in the cavity. M2 has a partially-reflectivity coating. The reflectivity of M2 is denoted by $R_{\rm M2}$. Thus, a portion of the rightward-traveling beam is allowed passing through M2. The transmissivities of lenses L1 and L2 are denoted by $\Gamma_{\rm L1}$ and $\Gamma_{\rm L2}$, respectively; namely, a little beam power in the cavity is eliminated by the lenses.  Here, M1 and the EOM can be combined as a reflective EOM, as M1 can be considered as a reflectivity coating attached to the EOM. We use $R_{\rm M1,EOM}$ to denote the reflectivity of this reflective EOM. Besides, the gain medium also has absorption/reflection. Therefore, we use $\Gamma_{\rm g}$ to represent its transmissivity. The air transmissivity is represented as $\Gamma_{\rm air}$. Furthermore, diffraction loss denoted by $\Gamma_{\rm diff}$ exists in the cavity. To calculate the output power, the SSLR is equivalent to a classical resonator which consists of a gain medium and two mirrors placed on the left and the right sides of the gain medium. The reflectivities of the equivalent mirrors are $\mathcal{R}_1=R_{\rm M1,EOM}\Gamma_{\rm L1}^2\Gamma_{\rm diff}\Gamma_{\rm g}^2$ and $\mathcal{R}_2=\Gamma_{\rm air}^2\Gamma_{\rm L2}^2R_{\rm M2}$, which count the  loss factors on the left side and the right of the gain medium of the SSLR, respectively. Then, the output power by M2 can be obtained as~\cite{a211015.06}
\begin{equation}
P_{\rm out}=\Gamma_{\rm M2}\Gamma_{\rm L2}\Gamma_{\rm air}\eta_{\rm slop}\left[ P_{\rm in} - P_{\rm th} \right],
\label{equ:P4reform}
\end{equation}
where
\begin{equation}
\eta_{\rm slop}=\dfrac{\eta_{\rm c} }{(1+\sqrt{\dfrac{\mathcal{R}_2}{\mathcal{R}_1}})(1- \sqrt{\mathcal{R}_1 \mathcal{R}_2})},
\label{equ:eta-slop}
\end{equation}
and
\begin{equation}
P_{\rm th}= \dfrac{\pi a_{\rm g}^2I_{\rm s}}{ \eta_{\rm c}}  \ln\dfrac{1}{\sqrt{\mathcal{R}_1\mathcal{R}_2}},
\label{equ:Pth}
\end{equation}
where $I_{\rm s}$ is the saturation intensity of the gain medium, and $\eta_{\rm c}$ is the combined efficiency in the pumping process. Note that the loss factors $\Gamma_{\rm g}$ and $\Gamma_{\rm diff}$ appear in the whole body of the gain medium, while for simplicity, we assemble these factors in the equivalent reflectivity $\mathcal{R}_1$. For most transmission distances, these losses are very low and therefore exhibit little effect on the result. If the diffraction loss is high, the results may deviate, but its trend remains the same as the distance increases.
The pump laser is generated by a laser diode and then absorbed by the gain medium. The efficiency from electricity to the absorbed power contains the pump diode's electro-optical conversion efficiency $\eta_{\rm p}$, the pump laser transmission efficiency $\eta_{\rm t}$, and the absorption efficiency $\eta_{\rm a}$. With respect to the internal action of the gain medium, there exist the quantum efficiency $\eta_{\rm Q}$, the Stokes factor $\eta_{\rm S}$, and the overlap efficiency $\eta_{\rm B}$. Thus, the combined efficiency is expressed as
\begin{equation}
\eta_{\rm c}=\eta_{\rm B}\eta_{\rm S} \eta_{\rm Q}\eta_{\rm a}\eta_{\rm t}\eta_{\rm p}.
\end{equation}

\subsection{Dynamic of Intra-cavity Resonant Beam}

As demonstrated in Fig.~\ref{fig:gain}(b), the photons travel back and forth circularly between two retroreflectors. During the circulation, the photons experience the optical amplification provided by the gain medium and also the attenuation induced by many loss factors. However, this process may bring instability to the resonant beam power. There are two aspects that affect the stability of the resonant beam power: the gain fluctuation and the loss fluctuation~\cite{a200508.02}. The absorption/reflection of lenses and the air as well as the transmission of mirrors can be static or dynamic. For example, if the receiver is moving, the counted reflectivity to the beam at each lens will change with respect to the incident angle. Besides, the EOM can be considered as a dynamic loss if intensity modulation is applied to the resonant beam. The intensity fluctuation also affects the gain medium, leading to the dynamic gain. Hence, we need to model the dynamic process of the resonant beam system and then create a simulation algorithm to observe the fluctuation of the beam intensity, especially when the beam is modulated. As a result of the existence of the echo signal,  the past signal is involved in modulation, and therefore, the output signal exhibits a chaos. Generally, it is difficult to demodulate such a chaotic signal. We do the time-domain simulation to find out a method for extracting  information from such a chaotic output.

We first study the dynamic behavior of the gain medium. As depicted in Fig.~\ref{fig:gain}(c), the atoms in the gain medium stay at different energy levels. For a four-level gain medium, such as Nd:YVO$_4$ crystal, the active  levels from low energy to high energy are denoted by E$_0$, E$_1$, E$_2$, and E$_3$, respectively. When the gain medium is irradiated by the pump laser with frequency $\nu_{\rm p}$,  atoms at E$_0$ transit to E$_3$, and then, quickly transit to a lower stable energy level E$_2$. There are two approaches for the excited atoms to transit from E$_2$ to E$_1$, namely, by spontaneous emission or by stimulated emission. Both of the above approaches emit photons with frequency $\nu_{\rm L}$. If there is no external photon with  frequency $\nu_{\rm L}$ passing through the gain medium, the excited atoms at E$_2$ will transit to lower energy levels spontaneously and meanwhile generate omnidirectional photons. This fluorescence decay operates with a slow rate, for example, the fluorescence time $\tau_{\rm f}$ of Nd:YVO$_4$ is $100~\mu$s. However, if the excited atom is stimulated by an input photon with frequency $\nu_{\rm L}$, it will transit to E$_1$ rapidly while generating an identical photon that has the same frequency and direction with the input one. Since the transitions E$_3$~$\to$~E$_2$ and E$_1$~$\to$~E$_0$ are very fast and nonradiative, few atoms remain at E$_3$ and E$_1$. Thus, we usually simplify the four-level system to a two-level system, as depicted in Fig.~\ref{fig:gain}(c).

The above dynamic process can be expressed by the following rate equations~\cite{a181218.01}:
\begin{align}
\label{equ:rate-equ1}	\displaystyle\frac{\partial N_2}{\partial t}&= - N_2\varphi\sigma c -\frac{N_2}{\tau_{\rm f}} + R_{\rm p},\\
\dfrac{\partial \varphi}{\partial t}&=  N_2\varphi\sigma c -\frac{ \varphi}{\tau_c} + S,
\label{equ:rate-equ2}
\end{align}
where $N_2$ is the atom density at E$_2$; $\varphi$ is the photon density in the cavity; $\sigma$ is the stimulated emission cross section of the gain medium; $c$ is the light speed; $\tau_{\rm f}$ is the fluorescence time; $R_{\rm p}$ is the pump rate; $\tau_c$ is the decay time of the intra-cavity photons; and $S$ is the rate at which spontaneous emission contributes to the stimulated emission. The pump rate and the spontaneous emission coupling rate are as follows:
\begin{equation}
R_{\rm p}=\dfrac{\eta_{\rm c} P_{\rm in}}{h\nu_{\rm L}V},~~\mbox{and}~~S=\beta\frac{ N_2}{\tau_{21}},
\end{equation}
where $h$ is the Planck's constant, $\nu_{\rm L}=c/\lambda$ is the photon frequency of the resonant beam, $V$ is the volume of the gain medium; $\beta$ is the spontaneous emission factor that represents the ratio of the spontaneous photons coupled into the laser resonator to the total generated spontaneous photons~\cite{a210918.01}; and $\tau_{21}$ is the decay time of the transition E$_2$~$\to$~E$_1$. Generally, $\beta$ is a very small number \mll{}(e.g., in this paper $\beta=0.1\%$)\mrr{}.
For computer simulation, the rate equations in~\eqref{equ:rate-equ1} and~\eqref{equ:rate-equ2} can be rewritten into the following discrete form:
\begin{align}
\displaystyle \Delta N_2&= (- N_2\varphi\sigma c -\frac{N_2}{\tau_{\rm f}} + R_{\rm p})\Delta t,\label{equ:rate-eq-disc1}\\
\Delta \varphi&=( N_2\varphi\sigma c -\frac{ \varphi}{\tau_{\rm c}} + S) \Delta t,\label{equ:rate-eq-disc2}
\end{align}
where $\Delta t$ represents the unit time in simulation; and $\Delta N_2$ and $\Delta \varphi$ are the  variation of the   excited atom density and the photon density per unit time, respectively. Every term in the equations represents a physical process. For instance, $-N_2\phi\sigma c$ describes the stimulated emission rate, $-N_2/\tau_{\rm f}$ is the decay rate of the excited atom density induced by  spontaneous emission, $R_{\rm p}$ is the increase rate of excited atom density produced by pumping process, and $\varphi/\tau_{\rm c}$ is the decay rate of intra-cavity photon density induced by cavity loss.

However, the above rate equations describe the behavior of the  average densities of the atoms and the photons and do not consider the fluctuation of these quantities within an oscillating period. Since the modulation period can be much shorter than the oscillating period, a micro perspective to the photon density variation is required. Here we set  $\Delta t= l_{\rm g}/c$, where $l_{\rm g}$ denotes the gain medium thickness. This means that a unit time of simulation is the time for the light to pass through the gain medium. \mll{}As depicted in Fig.~\ref{fig:gain}(b), the  densities of the rightward-traveling photons at the input and the output surfaces of the gain medium are $\varphi_1$ and $\varphi_2$, respectively. The densities of the leftward-traveling photons at the input and the output surfaces of the gain medium are $\varphi_3$ and $\varphi_4$, respectively.\mrr{} Then, from~\eqref{equ:rate-eq-disc1}, the density of the excited atoms is modeled as the following recursion formula:
\begin{equation}
\begin{array}{l}
N_2[n+1]= N_2[n]- N_2[n]\left\{\varphi_1[n]+\varphi_3[n]\right\}\sigma l_{\rm g} \vspace{1ex}\\~~~~~~~~~~~~~~-\dfrac{N_2 l_{\rm g}}{\tau_{\rm f} c} + \dfrac{R_{\rm p} l_{\rm g}}{c},
\end{array}\label{equ:sim-eq1}
\end{equation}
where $n$ is the time index. Equation~\eqref{equ:sim-eq1}
describes that the current atom density is derived from that before a time step. Similarly, the photon density can also be calculated by summing up the changes within a unit time. Here, we consider the photon densities at the two propagation directions. Then, we calculate the densities of the photons outputting from the gain medium by
\begin{align}
\varphi_2[n+1]=\varphi_1[n]+N_2[n]\varphi_1[n]\sigma l_{\rm g}+ \dfrac{S[n] l_{\rm g}}{2c}, \label{equ:sim-eq2}\\
\varphi_4[n+1]=\varphi_3[n]+N_2[n]\varphi_3[n]\sigma l_{\rm g}+ \dfrac{S[n] l_{\rm g}}{2c}.\label{equ:sim-eq3}
\end{align}
From~\eqref{equ:sim-eq2} and~\eqref{equ:sim-eq3}, we can see that the output photon densities from the gain medium are derived from the input photon densities. To complete the circulation, we build the relation between $\varphi_1$ and $\varphi_4$, and also the relation between $\varphi_3$ and $\varphi_2$; they are
\begin{align}
\varphi_1[n]&=\Gamma_{\rm g}^2\Gamma_{\rm diff}\Gamma_{\rm L1}^2R_{\rm M1,EOM}s[n-n_{\rm L}]\varphi_4[n-2n_{\rm L}],\\
\varphi_3[n]&=\Gamma_{\rm air}^2\Gamma_{\rm L2}^2R_{\rm M2}\varphi_2[n-2n_{\rm R}],
\label{equ:phi3-disc}
\end{align}
where $s[n]$ represents the modulation signal; and $n_{\rm L}$ and $n_{\rm R}$ are the time steps for the photons to transit from the gain medium to M1 and M2, respectively. Note that the time step of passing the gain medium is $1$, which should be considered in the subsequent analysis. The initial condition for the above equations is $\{N_2,\varphi_1,\varphi_2,\varphi_3,\varphi_4\}=0$.

\subsection{Modulation and Demodulation}

The information is modulated on the resonant beam by the EOM. Under this circumstance, we only consider using intensity modulation and direct detection~(IM/DD) scheme~\cite{imdd}. In this case, the EOM can be assumed as a multiplier, where the multiplier and the multiplicand are the input resonant beam intensity and the EOM's control signal, respectively. The EOM's control signal should be at least kept above a specific direct-current~(DC) level so that enough power is allowed to pass through the EOM to maintain the resonance. Therefore, given the source signal series $x[n]$ and the DC bias $p\in [0, 1]$, the control signal is expressed as
\begin{equation}
s[n]=(1-p)x[n]+p,
\end{equation}
where $(1-p)$ represents the signal amplitude, which ensures that the control signal is restricted within the range of $[0, 1]$. The beam power is counted as the total energy that passes the beam cross section within a unit time. As a photon with frequency $\nu_{\rm L}$ has the energy of $h\nu_{\rm L}$, and the beam cross section at the gain medium is approximate to be the cross section of the gain medium, the rightward-traveling beam power at the right-hand side of the gain medium yields
\begin{equation}
P_2[n]=\pi a_{\rm g}^2 c h\nu_{\rm L}\varphi_2[n].
\label{equ:phi2P2}
\end{equation}
Then, the output power $P_{\rm out}$ is obtained from $P_2$ by multiplying several loss factors; that is
\begin{equation}
P_{\rm out}[n]=\Gamma_{\rm M2}\Gamma_{\rm L2}\Gamma_{\rm air}P_2[n-n_{\rm R}],
\end{equation}
where $\Gamma_{\rm M2}=1-R_{\rm M2}$ is the transmissivity of M2.

According to the above analysis, it can be known that the current output signal is related to the past output signal. What is received by the receiver is the past signal that was reflected by the receiver and then experienced a series of transmission losses, the gain, and the intensity modulation in a circulation. According to this fact, we come out the demodulation method  relying on not only the present signal but also the past signal just before a circulating period $n_{\rm c}=2(n_{\rm L}+n_{\rm R}+1)$. This scheme can be expressed by a division, namely, the demodulated signal is obtained as
\begin{equation}
\begin{aligned}
y[n]&=\dfrac{P_{\rm out}[n]}{P_{\rm out}[n-n_{\rm c}]}\\
&=\Gamma_{\rm loss} G[n] s[n-n_{\rm c}/2],
\end{aligned}\label{equ:yn}
\end{equation}
and
\begin{equation}
\Gamma_{\rm loss}=R_{\rm M1,EOM}\Gamma_{\rm L1}^2\Gamma_{\rm g}^2 \Gamma_{\rm diff}\Gamma_{\rm air}^2\Gamma_{\rm L2}^2R_{\rm M2},
\label{equ:Gamma-loss}
\end{equation}
where $\Gamma_{\rm loss}$ contains all the static loss factors in the cavity, and $G[n]$ represents the dynamic gain of the gain medium. \mll{}$n_{\rm c}$ can be estimated by positioning technologies plus a signal processing method; we discuss this issue in the next Section.\mrr{} Note that we assume the loss is static, because its variation speed depends on the moving/rotating speed of receivers which is very slow compared with the modulation speed. For instance, the maximum moving speed of human is $44$~m/s~\cite{a180805.04}, we can derive that cutting off a beam with diameter of $1$~mm needs at least $22.7~\mu$s; this is very long in contrast to the modulation period which is below~$1$~ns.

\begin{figure}
	\centering
	\includegraphics[width=3.4in]{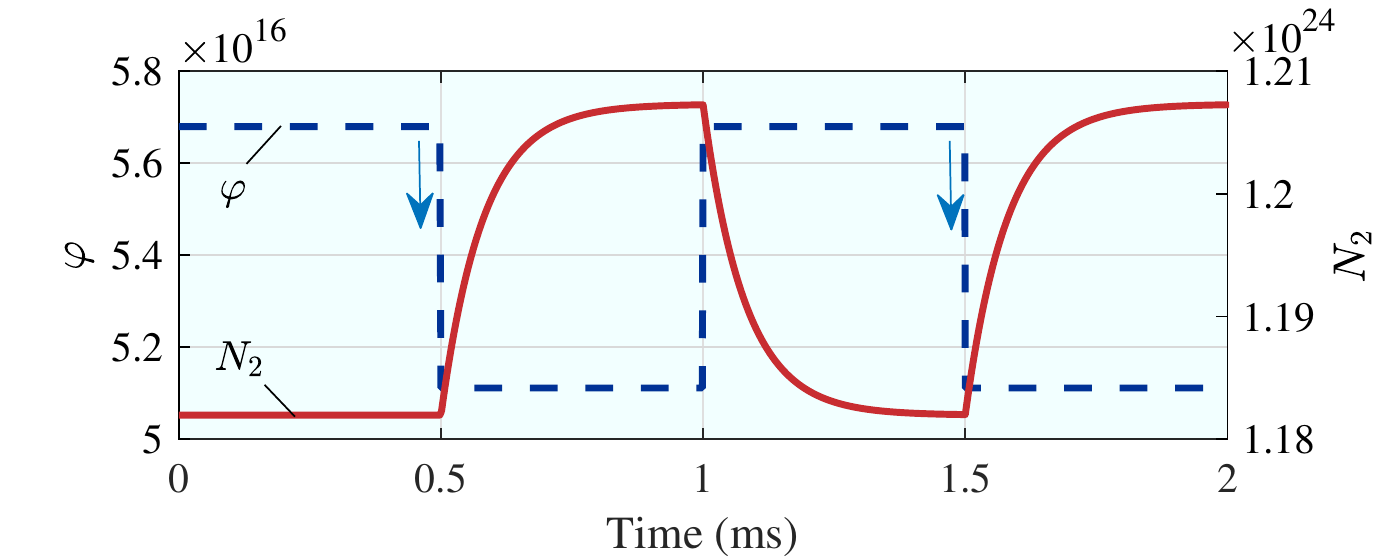}
	\caption{Response of the excited atom density to the change of  photon density}
	\label{fig:stepresp}
\end{figure}

From~\eqref{equ:yn}, we can see that the demodulated signal consists of three parts, i.e., the static loss, the dynamic gain, and the control signal. Consequently, if we want to retrieve the EOM's control signal correctly, the variation speed of $G[n]$ should be as slow as possible. Since the gain (related to the stimulated emission term $N_2\varphi\sigma c$) depends on the excited atom density $N_2$, we should ensure that the variation speed of $N_2$ is small compared with that of the photon density $\varphi$. Fortunately, this condition can be satisfied in the SSLR system under some specific conditions, as the gain medium can be viewed as a circuit with a capacitor (see Fig.~5 in~\cite{JZhou2021}) which stores  excited atoms and releases them with a relatively slow speed. For instance, assuming that the photon density in the gain medium is $\varphi=5.679339\times10^{16}$~m$^{-3}$ (i.e., the intra-cavity beam power is 10 W and the gain medium radius $a_{\rm g}=1$~mm) and the pump power is $20$~W, we can obtain that, at the stable state where $\partial N_2/ \partial t =0$, the excited atom density $N_2=1.181963\times10^{24}~$m$^{-3}$. Then, we assume a step change $\varphi_{\delta}$ is added to the photon density, i.e., $\varphi$ becomes $\varphi+\varphi_\delta$. We can see that the instantaneous change rate of the atom density becomes
\begin{equation}
\dfrac{\partial N_2}{\partial t}= -5.531587\times 10^{10}\times\varphi_\delta.
\end{equation}
If we set $\varphi_\delta=-0.1\varphi$ and the duration  $\Delta t=20$~ns (one period for photons to oscillate inside a $3$-m long resonator), we can obtain that the change of the atom density $\Delta N_2=6.283152\times10^{18}$, which is much less than $N_2$. Thus, we can ignore the effect induced by the variation of $N_2$. Intuitively, as demonstrated in Fig.~\ref{fig:stepresp}, we can observe that  the gain medium performs like a capacitor, and the step response time is almost at millisecond level, which is much longer than an information symbol. From the above analysis, we conclude that the gain of the gain medium $G[n]$ can be considered as a static value within a symbol duration under some specific parameter settings, and this is also verified by the  simulation results presented in Section~\ref{sec:evalu}.

\begin{figure*}
	\centering
	\includegraphics[width=6.6in]{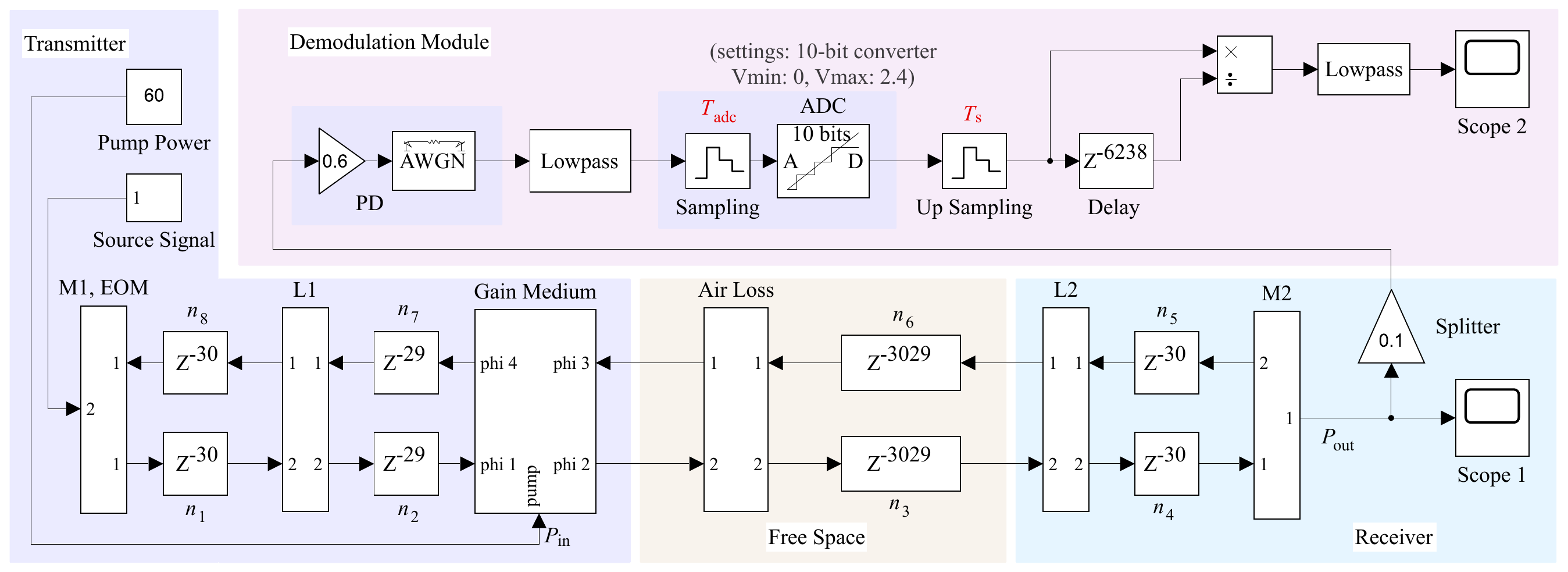}
	\caption{Simulation system implemented in the Simulink environment}
	\label{fig:simmain}
\end{figure*}

\section{Implementation}
\label{sec:imple}
As depicted in~(\ref{equ:sim-eq1})--(\ref{equ:phi3-disc}), the discrete rate equations provide the basic model for the time-domain simulation of the resonant beam system. In this section, we demonstrate using these equations to implement a simulation algorithm in  Matlab/Simulink software. In the next section, we use this simulation program to give an insight into the fluctuation of the intra-cavity resonant beam power at several stages, including initiating resonance, being interrupted, and rebuilding resonance, which are inevitable processes in the  resonant beam system. We also demonstrate how the intra-cavity echo interference arises when the resonant beam is modulated, and especially, verify the feasibility of the proposed delay-divide demodulation method.

\subsection{SSLR Simulation}
\begin{figure}[t]
	\centering
	\includegraphics[width=3.2in]{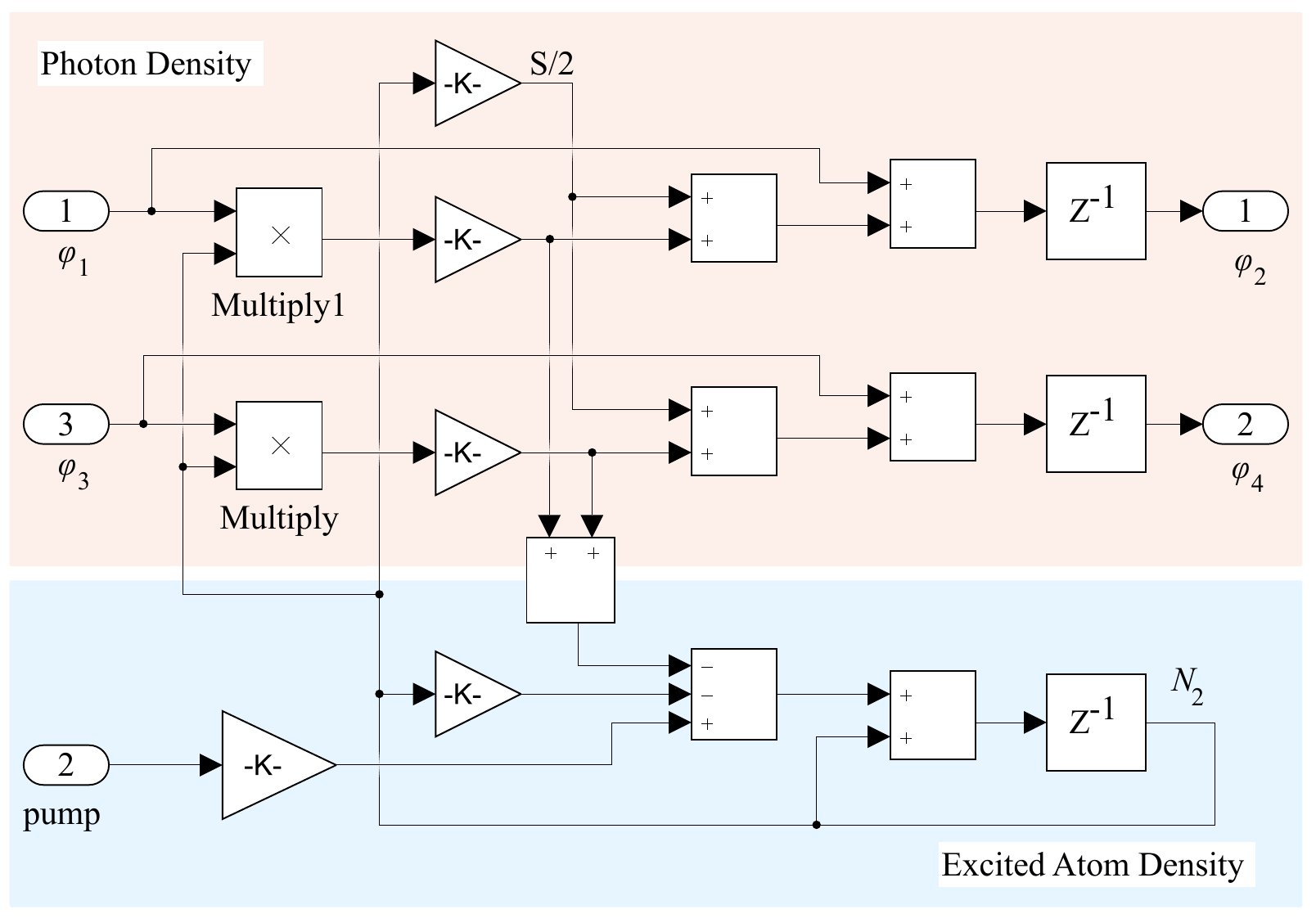}
	\caption{The gain medium model}
	\label{fig:simgain}
\end{figure}

As depicted in Fig.~\ref{fig:simmain}, the simulation system contains four parts, i.e., the transmitter, the free space, the receiver, and the demodulation module. Between any two optical devices, there is a delay model on each transmission line to model the beam transfer delay. Each optical device is represented by a customized Simulink model. Since this program only considers the intra-cavity power circulation without investigating the geometry of the beam, those passive elements such as the lenses, the mirrors, the air loss, and the splitter can be easily modeled by the gain model originally provided by Simulink. The most significant model created by us is the gain medium model which has four ports for the input and output of the traveling beams at oppose directions. The gain medium model simulates the dynamic behavior of the excited atom density $N_2$ and the photon density amplification as the beam passes, based on the discrete rate equations in (\ref{equ:sim-eq1})--(\ref{equ:sim-eq3}), as demonstrated in Fig.~\ref{fig:simgain}. The quantity in the circulation is the beam power, while the quantity computed in the gain medium model is the photon density; they have different units. Hence, a conversion/inversion between these two quantities should be added to the input/output ports of the gain medium model, using the relation expressed in~\eqref{equ:phi2P2}.

\begin{figure}[t]
	\centering
	\includegraphics[width=3.5in]{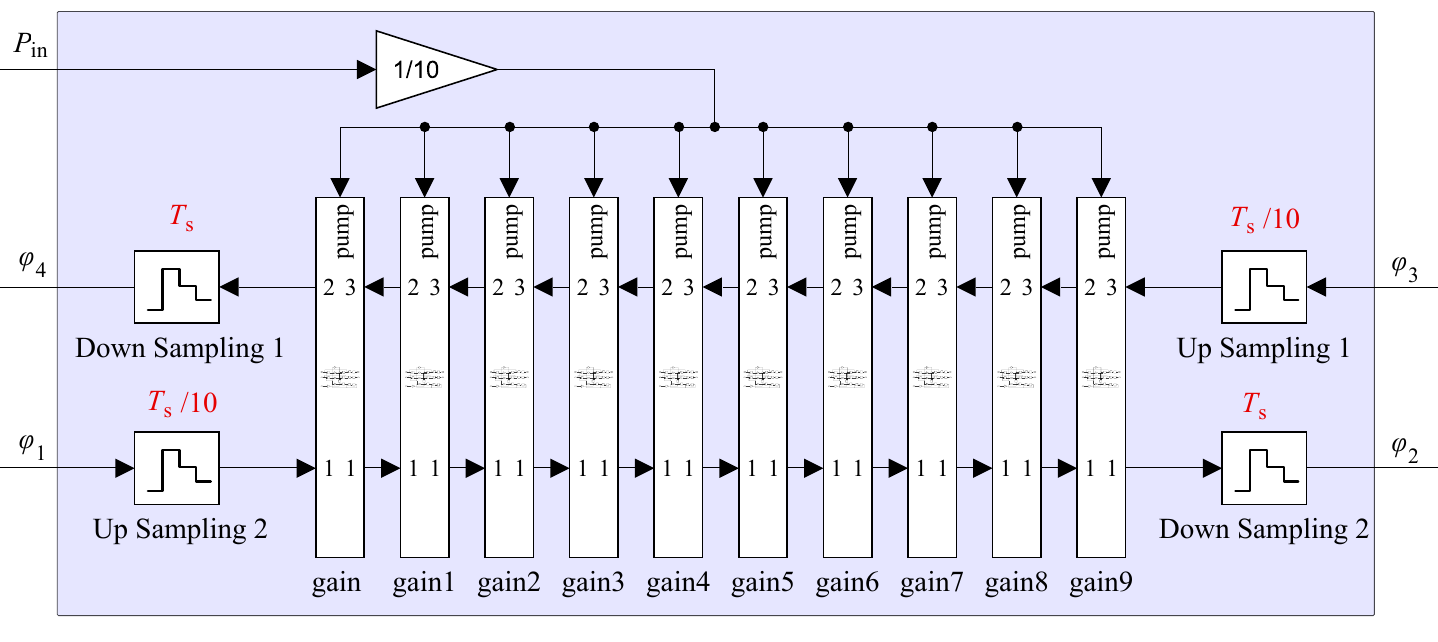}
	\caption{The gain medium model consists of several thinner slices}
	\label{fig:cascadegain}
\end{figure}

\begin{figure}[t]
	\centering
	\includegraphics[width=3.4in]{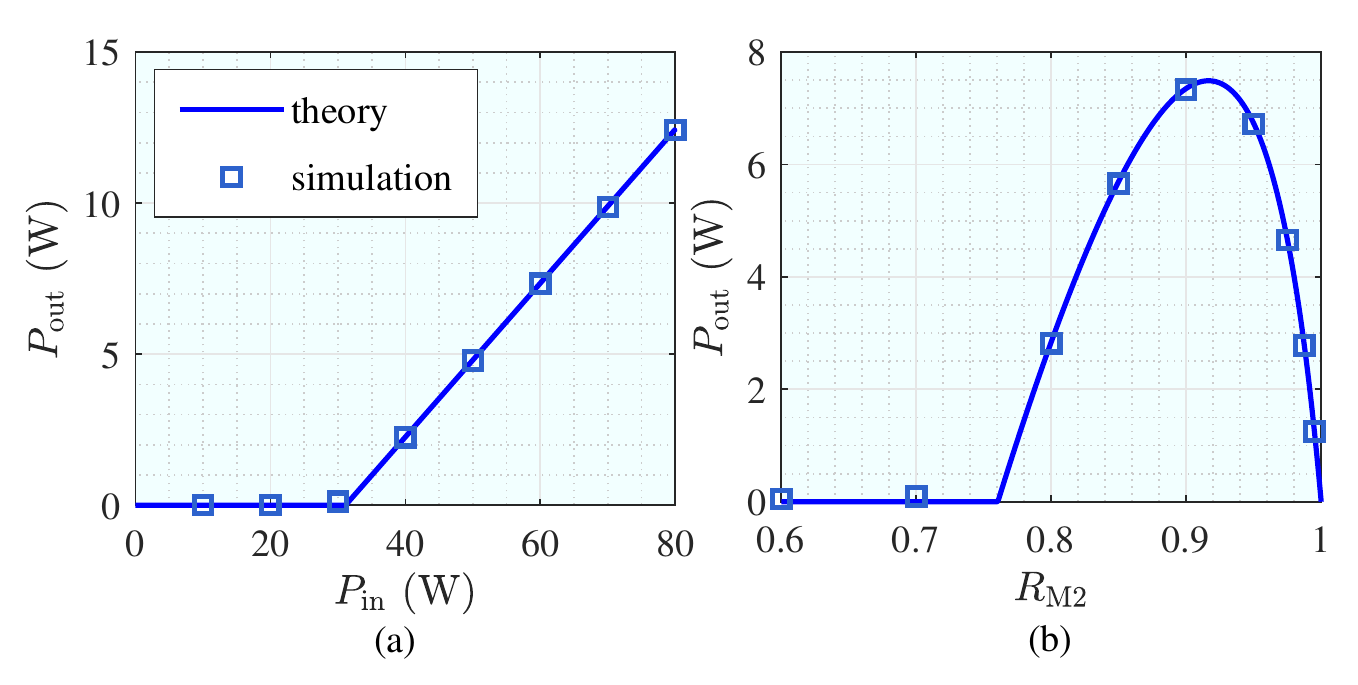}
	\caption{Comparison of stable-state output power $P_{\rm out}$ between simulation and theoretical computation under (a) different pump source power $P_{\rm in}$ and (b) different output mirror reflectivity $R_{\rm M2}$}
	\label{fig:cmpsimtheory}
\end{figure}

The simulation accuracy is determined by the length of time step and the optical gain of the gain medium. The discretization process may results in an incorrect decrease of the optical gain. To increase the accuracy, the gain medium is divided into 10 slices, as depicted in~Fig.~\ref{fig:cascadegain}. Here, each slice has the same implementation with the gain medium model depicted in Fig.~\ref{fig:simgain}. The pump source power $P_{\rm in}$ is also divided equally into 10 parts; and  each part is sent into a slice. This cascade model can well simulate the gradient of the gain value inside the gain medium, which greatly increases the simulation accuracy. Note that, before entering the cascade model, the input signal is upsampled with a faster period ($T_{\rm s}/10$) than the original sample period $T_{\rm s}$ to meet the time step requirement of each slice. After going out of the cascade model, the output signal is downsampled with the original period $T_{\rm s}$ to speed up the simulation. The parameters involved in simulation are listed in Table~\ref{tab:param}. Besides, we set $R_{\rm M1,EOM}=98.5\%$, $\{\Gamma_{\rm L1}, \Gamma_{\rm L2}, \Gamma_{\rm g}\}=99\%$. The air loss $\Gamma_{\rm air}$ and the diffraction loss $\Gamma_{\rm diff}$ are very small at $d=3$~m, which can be obtained according to~\cite{a200427.04} and \cite{a211015.06}, respectively. The output resonant beam power waveform can be observed by Scope 1. In this paper, unless otherwise specified, the pump source power $P_{\rm in}$ and the output mirror reflectivity $R_{\rm M2}$ are set to $60$~W and $90\%$, respectively. As shown in Fig.~\ref{fig:cmpsimtheory}, the stable-state output optical powers from the simulation results under different parameter settings are in good agreement with the theoretical results computed by the closed-form formula expressed in~(\ref{equ:P4reform})--(\ref{equ:Pth}).

\subsection{Delay-Divide Demodulation Module}
We also implement the delay-divide demodulation method in the simulation program. Here, we set the splitting ratio of optical power for demodulation is $r=0.1$. The PD's responsivity, which converts the optical power into current, is $\rho=0.6$~A/W. We assume the load resistor of the PD is $1~\Omega$, so that the output voltage is equal to the current. Both the splitter and the PD can be expressed by Simulink gain models. Since the EOM performs like a multiplier, the circulating signal inside the cavity will be moved to very-high frequency band by modulation process. However, the variation imposed on the intra-cavity signal within one round-trip does not exceed the frequency of the source signal. This condition gives us the opportunity of using a  lowpass filter~(LPF) to filter out the useless frequencies and meanwhile preserve the important information. Hence, the first LPF in the demodulation module is employed to remove the high-frequency components from the received signal. Next, the down sampling with period of $T_{\rm adc}$ models the sampling process of the analog-to-digital converter~(ADC), as the ADC model here operates continuously. The sample rate should be set large enough to prevent distortion of the passing signal, and we will investigate this issues in the next section. In Fig.~\ref{fig:simmain}, the ADC resolution is set to $10$ bits. The maximum input of the ADC is limited to be no greater than $2.5$~V.

After the conversion of ADC, the signal is stored in the memory as bits. Then, an upsampling model which can be realized by a digital processor in practice is placed on the line for accurate alignment. Next, a copy of the signal is sent to a delay model, which can be realized by a first input first output~(FIFO) memory.  The current signal is divided by the past signal received just before a delay time. This computes the variation of the intra-cavity loss and the gain within a round-trip, as verified by~\eqref{equ:yn}. Hence, the delay time equals  one period of circulation of the intra-cavity beam, i.e., it depends on the transmission distance. We can use positioning/ranging method to obtain the  distance in rough and then use the signal processor to find the accurate delay time. If the delay time is correct, the BER reaches the minimum value. The processed signal is then sent to the second LPF to remove any undesired frequencies generated from the division operation. Finally, on the output port of the second LPF, we can observe the demodulated signal by Scope 2. In the next section, we study the performance of the proposed  demodulation method.

	\section{Results}\label{sec:evalu}
	
	Now, we are able to use   time-domain simulations to observe the variation of the resonant beam power in a very small timescale under a very fast modulation frequency. We first observe the dynamic process when the resonance is initiated. Then, we observe the responses when foreign objects is inserted into the cavity and when the intra-cavity modulation is conducted. Next, we study the spectrum of the gain variation of the gain medium to clarify the reason of the question: Why does the proposed delay-divided demodulation scheme work? Finally, we obtain the BER of the proposed system and demonstrate the effects induced by the sample rate and the resolution of the ADC.

	\begin{table} [t]
		\caption{System  Parameters~\cite{a181218.01, a210622.02}}
		\renewcommand{\arraystretch}{1.2}
		\centering
		\begin{tabular}{ l l l}
			\hline
			\textbf{Parameter} & \textbf{Symbol} &  \textbf{Value} \\
			\hline
			Saturation intensity & $I_{\rm s}$ &  $1.1976\times10^7$~W/m$^2$\\
			Stimulated emission cross section & $\sigma$ & $15.6\times10^{-23}$ m$^{2}$\\
			Fluorescence lifetime&$\tau_{\rm f}$ & 100 $\mu$s\\
			Decay time from E$_2$ to E$_1$ & $\tau_{21}$& $115~\mu$s\\
			Resonant beam wavelength & $\lambda$ & $1064$ nm\\
			Combined pumping efficiency& $\eta_{\rm c}$ & $43.9\%$\\
			Spontaneous emission factor& $\beta$ & $0.001$\\
			Gain medium radius  & $a_{\rm g}$ & $2$ mm \\
			Gain medium thickness & $l_{\rm g}$ & $1$ mm\\
			Combined pumping efficiency& $\eta_{\rm c}$ & $43.9\%$\\
			Focal length of lens& $f$ & $30$ mm\\
			Mirror-to-lens interval& $l$& $30.225$ mm\\
			Transmission distance & $d$ &$3$ m \\
			DC bias & $p$ & $0.98$ V\\
			Splitting ratio & $r$ & $0.1$\\
			PD's responsivity& $\gamma$ & $0.6$~A/W\\
			\hline
		\end{tabular}
		\label{tab:param}
	\end{table}

	\subsection{Initiating Resonance }
	
\begin{figure}[t]
		\centering
		\textsf{}	\includegraphics[width=3.2in]{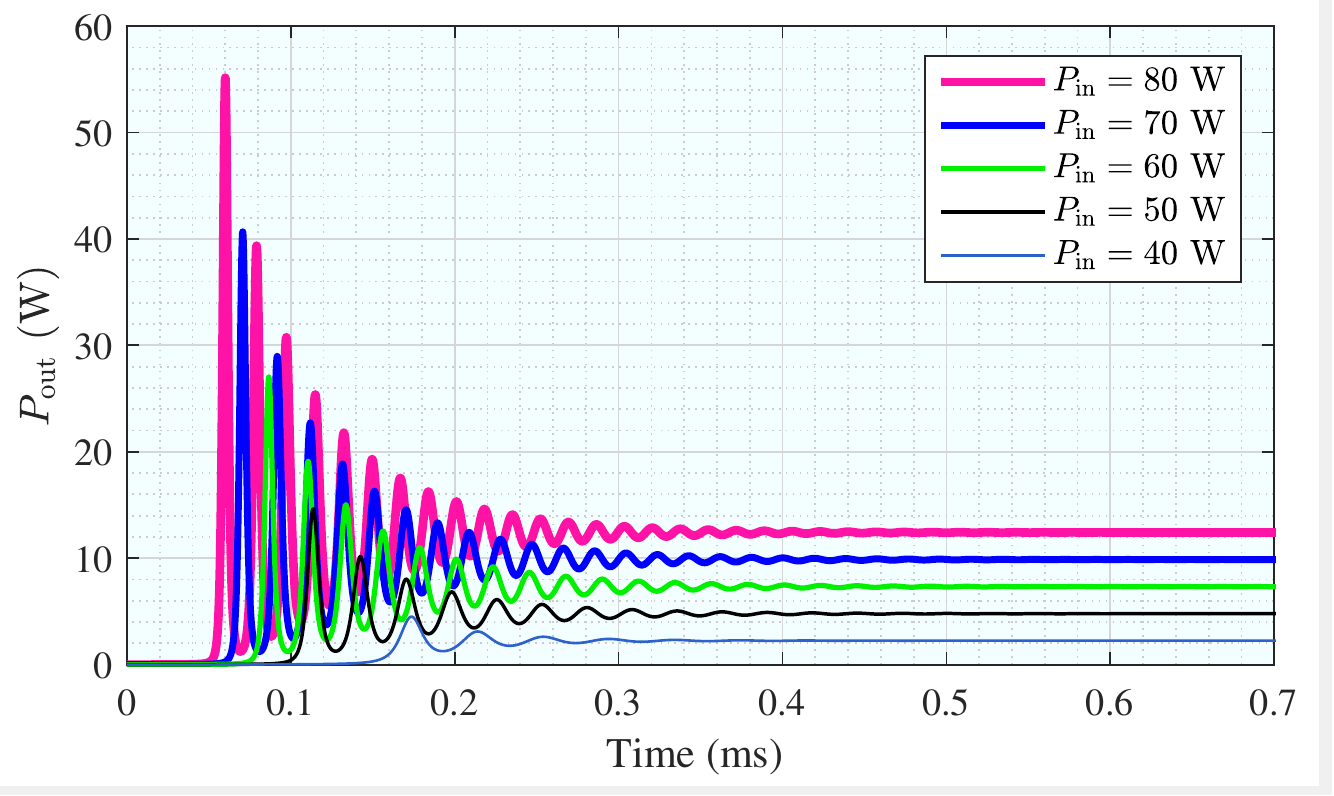}
		\caption{Relaxation oscillation under different pump source power $P_{\rm in}$}
		\label{fig:cmp-PoPin}
	\end{figure}

	 As the pump power starts to input the gain medium, the excited atom density increases rapidly, so the gain of the gain medium is very high. Then, the  resonant beam experiences a period of variation called \emph{relaxation oscillation}. As shown in Fig.~\ref{fig:cmp-PoPin}, at the beginning of the relaxation oscillation, the resonant beam power is quickly amplified to a very high level by stimulated emission. However, the excited atoms are also consumed during the stimulated emission process, and the decrease rate of the excited atom density raises with the increase of the resonant beam power. The resonant beam power  then starts to decrease and quickly drops to a very low level, since the excited atoms are largely consumed. The above oscillation   repeats  with a decay and reaches a stable state after several oscillation periods.

\begin{figure}[t]
		\centering
		\includegraphics[width=3.2in]{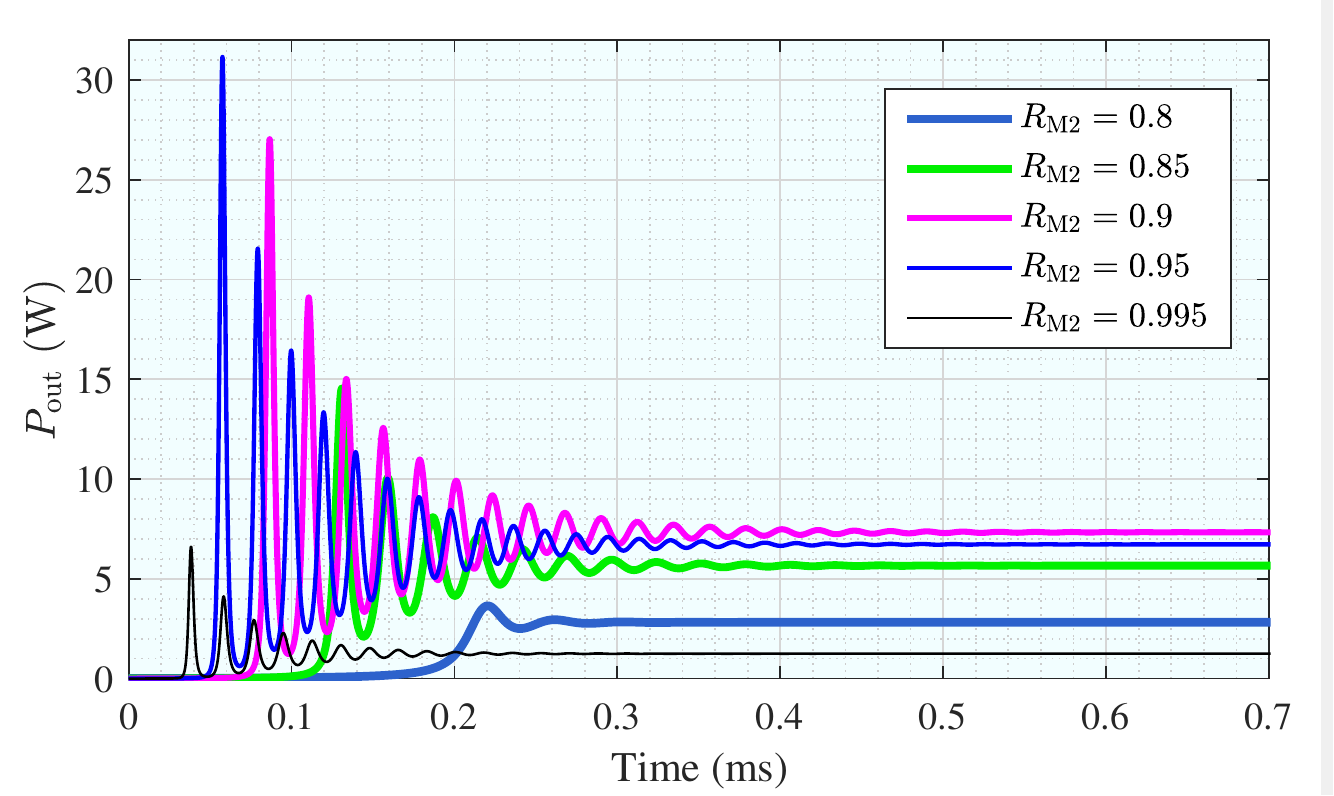}
		\caption{Relaxation oscillation under different output mirror reflectivity $R_{\rm M2}$}
		\label{fig:cmp-PoRm2}
	\end{figure}

	Fig.~\ref{fig:cmp-PoPin} depicts the output optical power $P_{\rm out}$ in time-domain under different pump power~$P_{\rm in}$. We can observe that with higher~$P_{\rm in}$,  the resonant beam experiences more severe relaxation oscillation. The oscillation frequency, the start-up time, and the peak power increases with the growth of~$P_{\rm in}$. Fig.~\ref{fig:cmp-PoRm2} demonstrates the relaxation oscillation under different output mirror reflectivity $R_{\rm M2}$. \mll{}We can see that higher $R_{\rm M2}$ leads to more violent relaxation oscillation.\mrr{} Although the peak output power under $R_{\rm M2}=0.995$ looks very low, the resonant beam power inside the cavity is very high, as most power is reflected back to the cavity by the mirror M2. From another perspective, we can compare the peak beam power with the beam power at the stable state and then find that under $R_{\rm M2}=0.995$ the relaxation oscillation is quite serious. We also find that the relaxation oscillation lasts for less than $0.5$~ms and the frequency is around $50$~kHz under the given parameter setting.
	
	\subsection{Response to Invasion and Modulation}
	
\begin{figure}[t]
		\centering
		\includegraphics[width=3.2in]{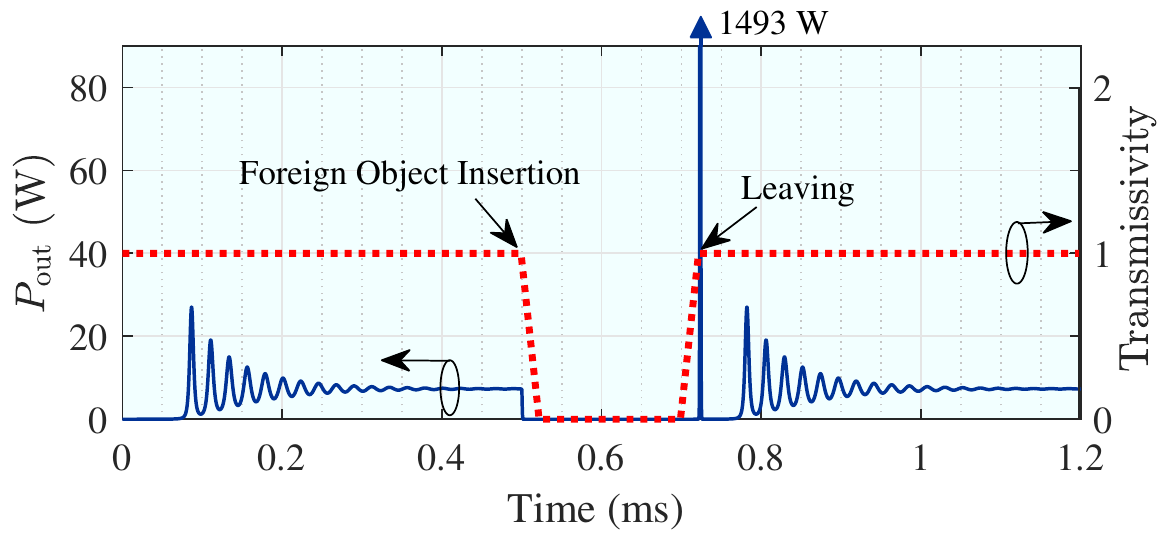}
		\caption{Output power response to  foreign object invasion}
		\label{fig:actresp-inv}
	\end{figure}

	Now, we study the response of the output beam power when an external influence is applied to the resonant beam. The most common case is that a foreign body is inserted into the cavity. The action of intrusion can be modeled as a continuing change of the intra-cavity diffraction loss. In simulation, we add a  transmissivity on the intra-cavity beam path to perform this action. The transmissivity decreases from $1$ to $0$ within $22.7~\mu$s (referring to the aforementioned maximum moving speed of human body) and then raises to $1$ again after a period of time. As shown in Fig.~\ref{fig:actresp-inv}, the resonance ceases quickly as the foreign object invades the cavity. As the object leaves, the resonance restarts with the occurrence of  relaxation oscillation. Different from the first initiation caused by the start-up of the pump source, the leaving of the foreign object causes a narrow pulse whose power is up to $1493$~W. This phenomenon arises from the retention of the excited atoms at the upper energy level $\mbox{E}_2$, as there is no photon contributing to the transition of the excited atoms when the beam path is obstructed. The optical gain provided by the gain medium becomes very high with the increase of the excited atom density. As the object leaves, the extremely high gain leads to the high-power pulse; and then, the excited atoms exhaust quickly, leading to the transient pulse duration. Because most of the excited atoms are consumed by the pulse, the subsequent relaxation oscillation pattern is similar to the first one. Since the pulse width is very small, it is unnecessary to concern  the safety of the receiving devices.
	
\begin{figure}[t]
	\centering
	\includegraphics[width=3.2in]{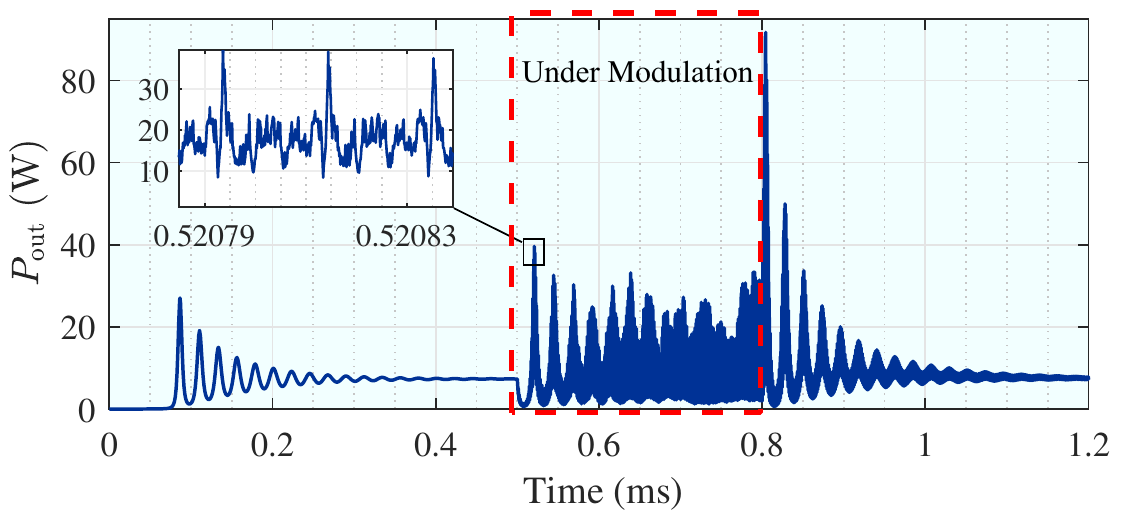}
	\caption{Output power response to  modulation and terminating modulation}
	\label{fig:actresp-mod}
\end{figure}
	
	In Fig~\ref{fig:actresp-mod}, we demonstrate the output power response to the modulation \mll{}(1-Gbit/s on-off keying signal with a DC bias)\mrr{} conducted by the EOM. It can be observed that the beam power starts to fluctuate as the modulation is conducted. It can be found that the variation of the output power is the superposition of a fast fluctuation and a slow fluctuation (see the envelope). The fast fluctuation is caused by the modulation, although we cannot directly recognize the source signal  from the wave. The slow fluctuation comes from the lose of balance of the resonant system which happens when the loss or the input power is changed. Here, the change of the total loss is induced by modulation, since the EOM changes its transmissivity during  modulation. As the modulation stops, we can also observe a short pulse which is similar to Fig.~\ref{fig:actresp-inv}. This phenomenon has the same reason with the leaving process of a foreign object, as the termination of the modulation reduces the intra-cavity loss suddenly. The subsequent process is also similar to Fig.~\ref{fig:actresp-inv}, but we can see that the fast fluctuation continues. The fast fluctuation after terminating the modulation comes from the circular transfer of the intra-cavity beam (the reason of echo interference). Nevertheless, without modulation, the fast fluctuation exhibits a decay as time goes on.

	\subsection{Demodulation}
\begin{figure}
	\centering
	\includegraphics[width=3.2in]{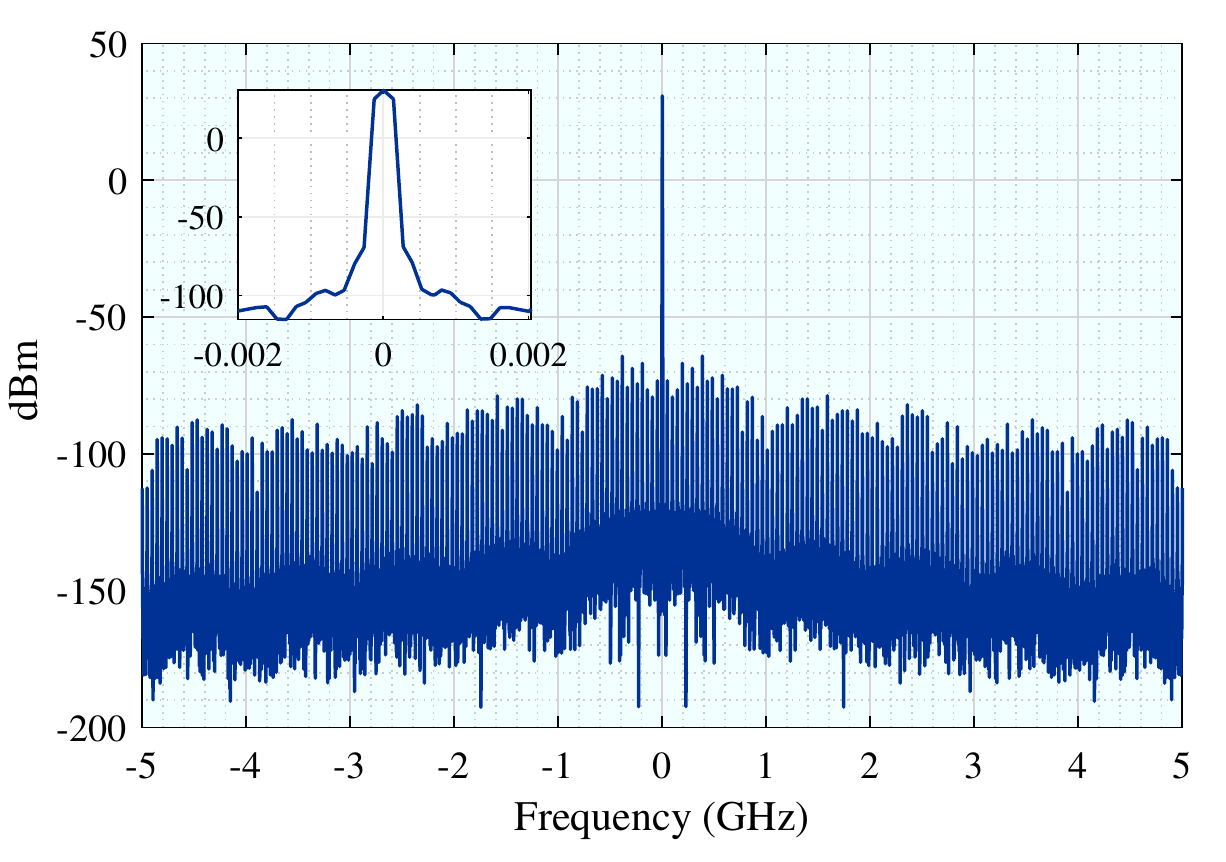}
	\caption{Spectrum of the gain fluctuation of the gain medium}
	\label{fig:gainspectrum}
\end{figure}

\begin{figure}[t]
		\centering
		\includegraphics[width=3.5in]{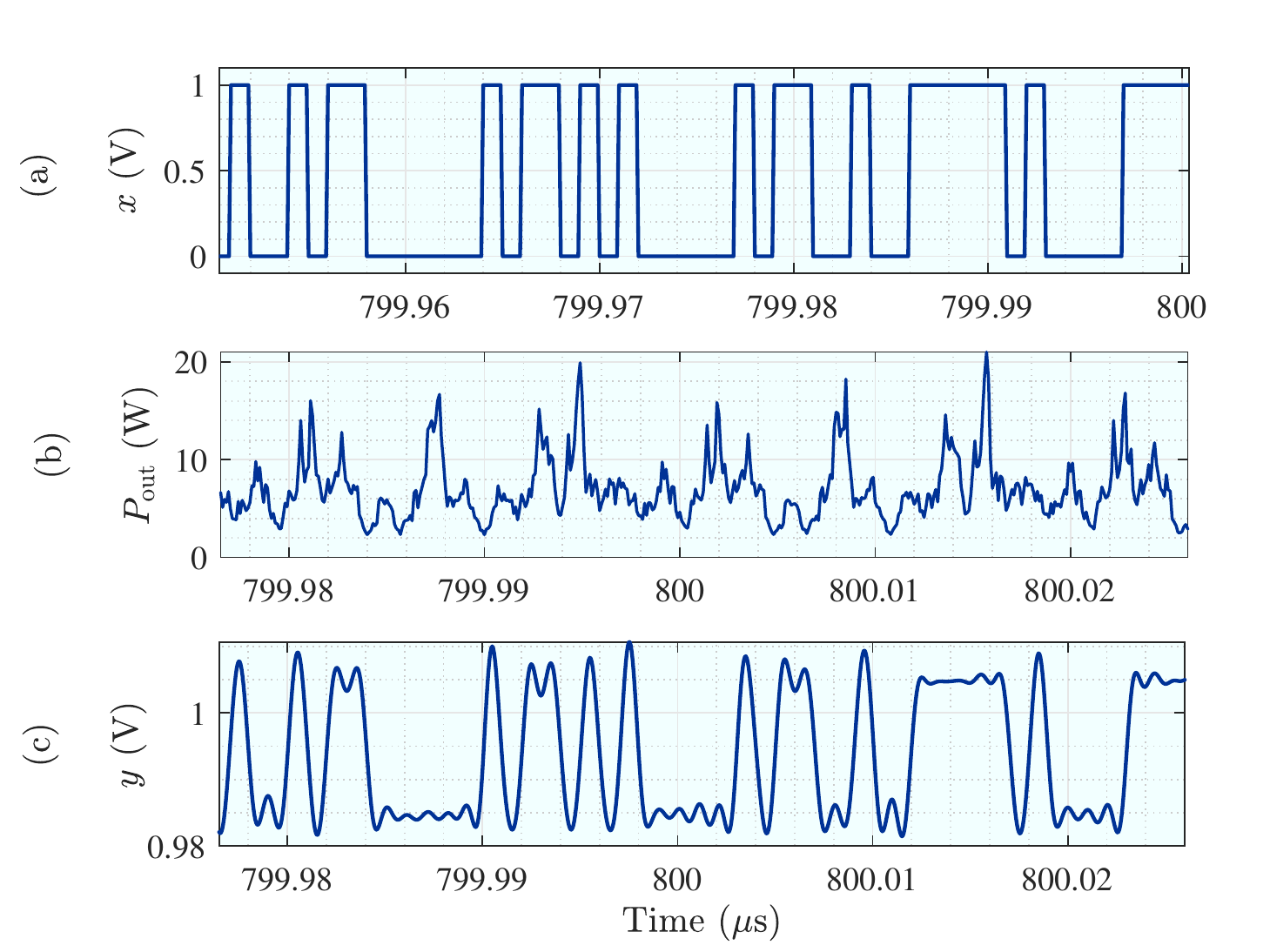}
		\caption{Demonstration of demodulation: (a) the source signal; (b) the output signal at mirror M2; and (c) the demodulated signal}
		\label{fig:moduldemodul}
	\end{figure}

	From Fig.~\ref{fig:actresp-mod}, we can find that it is quite difficult to retrieve the source information from such a chaotic output. Fortunately, the slow variation speed of the gain of the gain medium provides the opportunity of demodulating the signal. We have put forward the delay-divide demodulation method in Section~II.C. Now, we exploit simulation to observe the spectrum of the gain variation and then find if the requirement of demodulation can be met. The waveform of the gain is obtained by computing the ratio of the output power from the gain medium model to the input which was recorded (using a delay model) just when the photon entering the gain medium. We calculated the spectrum of the gain waveform, as depicted in Fig.~\ref{fig:gainspectrum}. It can be found that the significant frequency of the gain waveform is less than $250$~kHz, which is much smaller than the modulation frequency. As demonstrated in~\eqref{equ:yn}, the demodulation is  just realized by computing the round-trip change (induced by the gain and the total loss) of the intra-cavity resonant beam power. So far, we have found that both the gain and the loss can be viewed as static values in small timescale. The slow change of the gain in large timescale can be overcome by using training sequence which is a common method in wireless communications. Fig.~\ref{fig:moduldemodul} demonstrates an example of modulation and demodulation. It is easy to find that the demodulated signal in Fig.~\ref{fig:moduldemodul}(c)  matches the source signal in Fig.~\ref{fig:moduldemodul}(a) well, although the variation of the output power in Fig.~\ref{fig:moduldemodul}(b) looks chaotic.


	\subsection{Bit Error Rate}
\begin{figure}[t]
	\centering
	\includegraphics[width=3.4in]{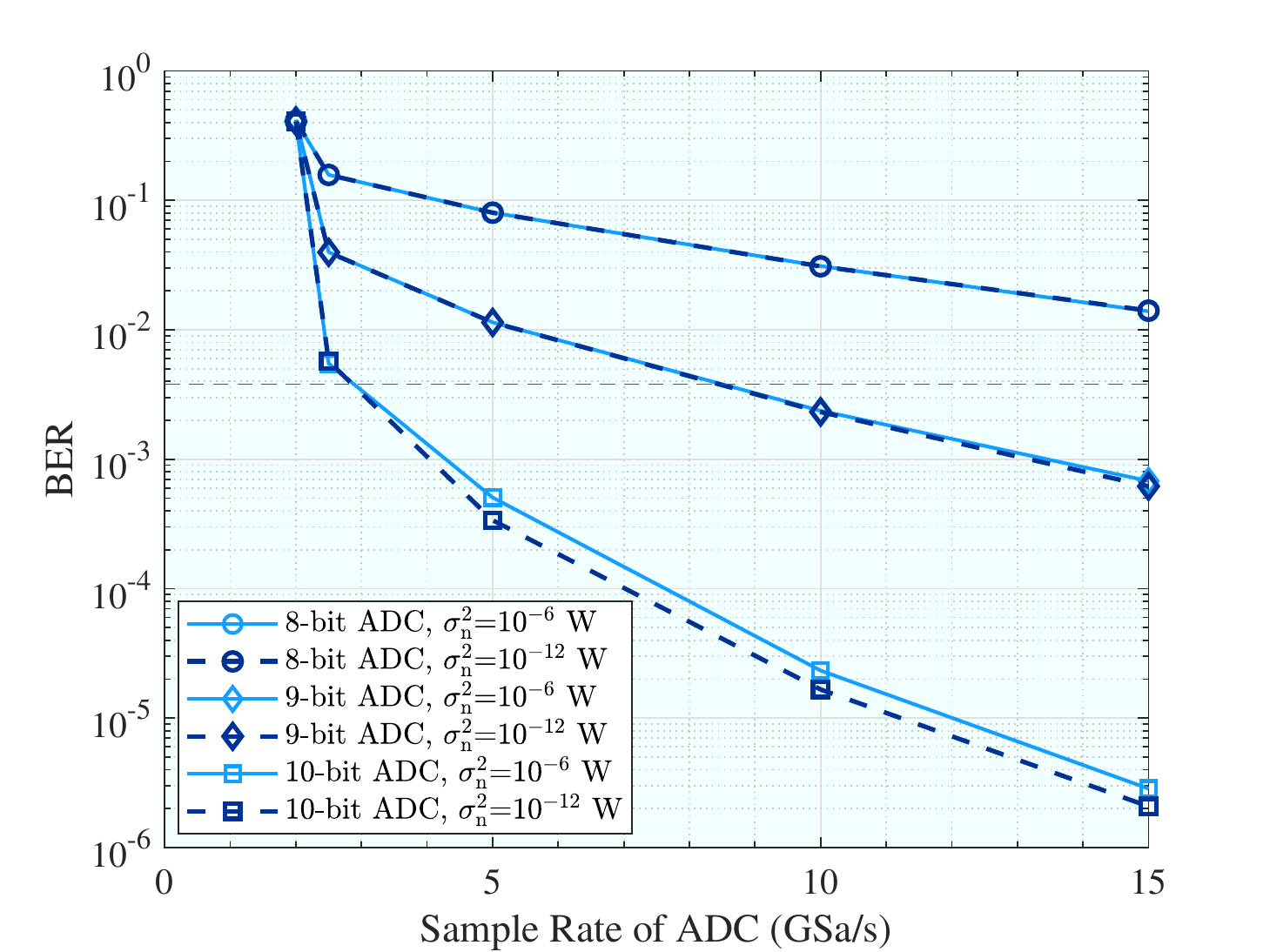}
	\caption{Bit error rate \emph{vs.}  ADC sample rate under different ADC resolution bits and noise variance $\sigma_{\rm n}^2$}
	\label{fig:ber}
\end{figure}

	We conduct a communication simulation to evaluate the performance of the proposed demodulation method. The source signal is $1$-Gbit/s binary data. Thus, the passband edge frequency and the stopband edge frequency of each LPF are set to $1$~GHz and $1.2$~GHz, respectively. The source signal is added to a DC bias, forming the control signal of the EOM. The bias $p=0.98$~V, and thus, the amplitude of source is limited to be $0.02$~V. We assume the conversion from the control signal to the EOM's transmissivity is not scaled. The pump power $P_{\rm in}$ is set to $60$~W. Referring to Fig.~\ref{fig:cmpsimtheory}(b), we choose $R_{\rm M2}=90\%$ to obtain a high output power as possible. The modulation is initiated $0.5$~ms after the pump source being turned on. The demodulated waveform is then sampled with the same rate of the source symbol, but a phase offset is applied to this sampling operation to meet the best sampling position. These samples is split into several segments, and each segment contains $500$ samples. These segments are processed independently to overcome the slow fluctuation of the gain. The zero-one  determination threshold for the samples in a segment is the average value of these samples. BER is counted for evaluating the communication performance. We examine several factors that may affect the  performance, including the ADC's sample rate, the ADC's resolution bits, and the noise level. As demonstrated in Fig.~\ref{fig:ber}, with the increase of the ADC's sample rate, the BER decreases relatively. ADC with 10-bit resolution provides better performance than that with 8-bit resolution. Due to the high power of the receiving signal, the noise exhibits less effect on the performance. We can see that most simulation cases can achieve BER $>3.8\times 10^{-3}$ (the threshold of 7\% hard-decision forward error correction~\cite{a190909.54}) when the ADC has $10$-bit resolution. This result shows that it is feasible to extract the source signal directly from the waveform of the resonant beam power without any complex modulation schemes and redundant optical devices.

	\section{Conclusions}
	\label{sec:con}
	In this paper, a time-domain simulation algorithm for the resonant beam charging and communication system and a delay-divide demodulation method are proposed. The simulation algorithm can monitor the variation of every dynamic parameter in the resonant beam system with a very small timescale which is even less than the time for photons to pass through an optical device. Also, the system response to the fast modulation applied to the intra-cavity beam can also be monitored by the simulation algorithm. We compared the simulation results at the final stable state with the theoretical computation, verifying the high accuracy of the simulation algorithm. Using the simulation algorithm, we observed the system response to four important actions, including the start-up of pumping, the invasion and  leaving of foreign objects, and the intra-cavity modulation.  By observing the spectrum of the gain fluctuation waveform, we found that the frequency of the gain fluctuation is much smaller than the modulation frequency, verifying that the required condition of the proposed demodulation method is satisfied. The  demodulation method is also realized in the simulation system. We  conducted on-off keying modulation and count the bit error rate to demonstrate the communication performance, which also verify the feasibility of extracting the source information directly from the chaotic waveform of the received resonant beam power.


	
	%

	%



	\ifCLASSOPTIONcaptionsoff
	\newpage
	\fi

	
	
	
	\bibliographystyle{IEEETran}
	\small
	%
	\bibliography{mybib}

\begin{thebibliography}{10}
\providecommand{\url}[1]{#1}
\csname url@samestyle\endcsname
\providecommand{\newblock}{\relax}
\providecommand{\bibinfo}[2]{#2}
\providecommand{\BIBentrySTDinterwordspacing}{\spaceskip=0pt\relax}
\providecommand{\BIBentryALTinterwordstretchfactor}{4}
\providecommand{\BIBentryALTinterwordspacing}{\spaceskip=\fontdimen2\font plus
\BIBentryALTinterwordstretchfactor\fontdimen3\font minus
  \fontdimen4\font\relax}
\providecommand{\BIBforeignlanguage}[2]{{%
\expandafter\ifx\csname l@#1\endcsname\relax
\typeout{** WARNING: IEEEtran.bst: No hyphenation pattern has been}%
\typeout{** loaded for the language `#1'. Using the pattern for}%
\typeout{** the default language instead.}%
\else
\language=\csname l@#1\endcsname
\fi
#2}}
\providecommand{\BIBdecl}{\relax}
\BIBdecl

\bibitem{a211015.01}
J.~D.~N. Dionisio, W.~G. Burns~III, and R.~Gilbert, ``{3D} virtual worlds and
  the metaverse: Current status and future possibilities,'' \emph{ACM Comput.
  Surv.}, vol.~45, no.~3, pp. 1--38, July 2013.

\bibitem{a211015.03}
Z.~Fei, F.~Wang, J.~Wang, and X.~Xie, ``{QoE} evaluation methods for 360-degree
  {VR} video transmission,'' \emph{IEEE J. Sel. Topics Signal Process.},
  vol.~14, no.~1, pp. 78--88, Jan. 2020.

\bibitem{a211015.02}
K.~Oiwake, K.~Komiya, H.~Akasaki, and T.~Nakajima, ``{VR} classroom: Enhancing
  learning experience with virtual class rooms,'' in \emph{Eleventh
  International Conference on Mobile Computing and Ubiquitous Network (ICMU)},
  Auckland, New Zealand, Oct. 2018, pp. 1--6.

\bibitem{a180820.09}
K.~David and H.~Berndt, ``{6G} vision and requirements: Is there any need for
  beyond {5G}?'' \emph{IEEE Veh. Technol. Mag.}, vol.~13, no.~3, pp. 72--80,
  July 2018.

\bibitem{a180727.01}
Q.~{Liu}, J.~{Wu}, P.~{Xia}, S.~{Zhao}, W.~{Chen}, Y.~{Yang}, and L.~{Hanzo},
  ``Charging unplugged: Will distributed laser charging for mobile wireless
  power transfer work?'' \emph{IEEE Veh. Technol. Mag.}, vol.~11, no.~4, pp.
  36--45, Nov. 2016.

\bibitem{a200528.01}
M.~{Xiong}, Q.~{Liu}, M.~{Liu}, X.~{Wang}, and H.~{Deng}, ``Resonant beam
  communications with photovoltaic receiver for optical data and power
  transfer,'' \emph{IEEE Trans. Commun.}, vol.~68, no.~5, pp. 3033--3041, May
  2020.

\bibitem{a200508.02}
M.~{Xiong}, Q.~{Liu}, G.~{Wang}, G.~B. {Giannakis}, and C.~{Huang}, ``Resonant
  beam communications: Principles and designs,'' \emph{IEEE Commun. Mag.},
  vol.~57, no.~10, pp. 34--39, Oct. 2019.

\bibitem{a180805.04}
V.~Iyer, E.~Bayati, R.~Nandakumar, A.~Majumdar, and S.~Gollakota, ``Charging a
  smartphone across a room using lasers,'' \emph{Proc. ACM Interact. Mob.
  Wearable Ubiquitous Technol.}, vol.~1, no.~4, pp. 1--21, Jan. 2017.

\bibitem{a191111.01}
T.~Koonen, A.~Khalid, J.~Oh, F.~Gomez-Agis, and E.~Tangdiongga, ``High-capacity
  optical wireless communication using 2-dimensional {IR} beam steering,'' in
  \emph{Opto-Electronics and Communications Conference (OECC) and Photonics
  Global Conference (PGC)}, Singapore, Nov. 2017, pp. 1--4.

\bibitem{a190611.03}
T.~Koonen, J.~Oh, K.~Mekonnen, and E.~Tangdiongga, ``Ultra-high capacity indoor
  optical wireless communication using steered pencil beams,'' in \emph{Proc.
  International Topical Meeting on Microwave Photonics}, Paphos, Cyprus, Oct.
  2015, pp. 4802--4809.

\bibitem{a201201.02}
H.~{Rhee}, J.~{You}, H.~{Yoon}, K.~{Han}, M.~{Kim}, B.~G. {Lee}, S.~{Kim}, and
  H.~{Park}, ``32 {Gbps} data transmission with {2D} beam-steering using a
  silicon optical phased array,'' \emph{IEEE Photon. Technol. Lett.}, vol.~32,
  no.~13, pp. 803--806, May 2020.

\bibitem{a190514.01}
Z.~{Zhang}, J.~{Dang}, L.~{Wu}, H.~{Wang}, J.~{Xia}, W.~{Lei}, J.~{Wang}, and
  X.~{You}, ``Optical mobile communications: Principles, implementation, and
  performance analysis,'' \emph{IEEE Trans. Vehi. Technol.}, vol.~68, no.~1,
  pp. 471--482, Nov. 2019.

\bibitem{a210901.06}
C.~E. O'Lone, H.~S. Dhillon, and R.~Michael~Buehrer, ``Characterizing the
  first-arriving multipath component in {5G} millimeter wave networks: {TOA},
  {AOA}, and non-line-of-sight bias,'' \emph{IEEE Trans. Wireless Commun.},
  2021, to appear, doi:10.1109/TWC.2021.3105641.

\bibitem{a190318.02}
G.~J. Linford, E.~R. Peressini, W.~R. Sooy, and M.~L. Spaeth, ``Very long
  lasers,'' \emph{Appl. Opt.}, vol.~13, no.~2, pp. 379--390, Feb. 1974.

\bibitem{a190318.01}
G.~J. Linford and L.~W. Hill, ``{Nd:YAG} long lasers,'' \emph{Appl. Opt.},
  vol.~13, no.~6, pp. 1387--1394, June 1974.

\bibitem{wicharge}
R.~Della-Pergola, O.~Alpert, O.~NAHMIAS, and V.~Vaisleib, ``Spatially
  distributed laser resonator,'' Worldwide Patent WO2\,012\,172\,541A1, Dec.
  20, 2012.

\bibitem{a211015.06}
M.~Xiong, M.~Liu, Q.~Jiang, J.~Zhou, Q.~Liu, and H.~Deng, ``Retro-reflective
  beam communications with spatially separated laser resonator,'' \emph{IEEE
  Trans. Wireless Commun.}, vol.~20, no.~8, pp. 4917--4928, Aug. 2021.

\bibitem{liu2021mobile}
Q.~Liu, M.~Xiong, M.~Liu, Q.~Jiang, W.~Fang, and Y.~Bai, ``Mobile wireless
  power transfer using a self-aligned resonant beam,'' \emph{arXiv preprint
  arXiv:2105.13174}, 2021.

\bibitem{a210831.01}
W.~Wang, Y.~Gao, D.~Sun, X.~Du, J.~Guo, and X.~Liang, ``Adjustable-free and
  movable {Nd:YVO$_4$} thin disk laser based on the telecentric cat’s eye
  cavity,'' \emph{Chin. Opt. Lett.}, vol.~19, no.~11, p. 111403, Aug. 2021.

\bibitem{a211015.04}
W.~Fang, H.~Deng, Q.~Liu, M.~Liu, Q.~Jiang, L.~Yang, and G.~B. Giannakis,
  ``Safety analysis of long-range and high-power wireless power transfer using
  resonant beam,'' \emph{IEEE Trans. Signal Process.}, vol.~69, pp. 2833--2843,
  May 2021.

\bibitem{a210823.02}
M.~Liu, M.~Xiong, Q.~Liu, S.~Zhou, and H.~Deng, ``Mobility-enhanced
  simultaneous lightwave information and power transfer,'' \emph{IEEE Trans.
  Wireless Commun.}, vol.~20, no.~10, Oct. 2021.

\bibitem{a190926.02}
J.~Lim, T.~S. Khwaja, and J.~Ha, ``Wireless optical power transfer system by
  spatial wavelength division and distributed laser cavity resonance,''
  \emph{Opt. Express}, vol.~27, no.~12, pp. A924--A935, June 2019.

\bibitem{a210901.01}
D.~Li, Y.~Tian, and C.~Huang, ``Capacity analysis of mobile resonant beam
  communications,'' in \emph{IEEE International Conference on Communications
  (ICC)}, 2021, to appear, doi:10.1109/ICC42927.2021.9500944.

\bibitem{a211015.07}
M.~Xiong, Q.~Liu, G.~Wang, G.~B. Giannakis, S.~Zhang, J.~Zhu, and C.~Huang,
  ``Resonant beam communications with echo interference elimination,''
  \emph{IEEE Internet Things J.}, vol.~8, no.~4, pp. 2875--2885, Feb. 2021.

\bibitem{MXiong2021.InSHG}
M.~Xiong, Q.~Liu, X.~Wang, S.~Zhou, B.~Zhou, and Z.~Bu, ``Mobile optical
  communications using second harmonic of intra-cavity laser,'' \emph{arXiv
  preprint arXiv:2106.11116}, 2021.

\bibitem{a210706.01}
R.~Mys, ``Time dependent simulation of laser by {Gauss-modes},'' Master's
  thesis, Friedrich Alexander University Erlangen-Nuremberg, Sept. 2005.

\bibitem{a181220.01}
T.~Schaer, R.~Rusnov, S.~Eagle, J.~Jastrebski, S.~Albanese, and X.~Fernando,
  ``A dynamic simulation model for semiconductor laser diodes,'' in
  \emph{Canadian Conference on Electrical and Computer Engineering. Toward a
  Caring and Humane Technology (Cat. No.03CH37436)}, May 2003, pp. 293--297.

\bibitem{JZhou2021}
J.~Zhou, M.~Xiong, M.~Liu, Q.~Liu, and S.~Zhou, ``Transient analysis for
  resonant beam charging and communication,'' \emph{IEEE Internet Things J.},
  2021, to appear, doi:10.1109/JIOT.2021.3094809.

\bibitem{a200522.04}
V.~Magni, ``Multielement stable resonators containing a variable lens,''
  \emph{J. Opt. Soc. Am. A}, vol.~4, no.~10, pp. 1962--1969, Oct. 1987.

\bibitem{a181221.01}
N.~Hodgson and H.~Weber, \emph{Laser Resonators and Beam Propagation:
  Fundamentals, Advanced Concepts and Applications 2nd ed.}\hskip 1em plus
  0.5em minus 0.4em\relax New York, NY., U.S.: Springer, 2005.

\bibitem{a181218.01}
W.~Koechner, \emph{Solid-State Laser Engineering, 6th ed.}\hskip 1em plus 0.5em
  minus 0.4em\relax New York, NY, USA: Springer, 2006.

\bibitem{a210918.01}
M.~Van~Exter, G.~Nienhuis, and J.~Woerdman, ``Two simple expressions for the
  spontaneous emission factor $\beta$,'' \emph{Phys. Rev., A}, vol.~54, no.~4,
  p. 3553, Oct. 1996.

\bibitem{imdd}
L.~Chen, B.~Krongold, and J.~Evans, ``Theoretical characterization of nonlinear
  clipping effects in {IM/DD} optical {OFDM} systems,'' \emph{IEEE Trans.
  Communi.}, vol.~60, no.~8, pp. 2304--2312, Aug. 2012.

\bibitem{a200427.04}
I.~I. Kim, B.~Mcarthur, and E.~J. Korevaar, ``Comparison of laser beam
  propagation at 785 nm and 1550 nm in fog and haze for optical wireless
  communications,'' \emph{in Proc. SPIE}, vol. 4214, no.~2, pp. 26--37, Feb.
  2001.

\bibitem{a210622.02}
W.~Long, T.~Wu, J.~Jiao, M.~Tang, and M.~Xu, ``Refraction-learning-based whale
  optimization algorithm for high-dimensional problems and parameter estimation
  of {PV} model,'' \emph{Eng. Appl. Artif. Intell.}, vol.~89, p. 103457, Mar.
  2020.

\bibitem{a190909.54}
L.-Y. Wei, C.-W. Hsu, C.-W. Chow, and C.-H. Yeh, ``20.231 {Gbit/s} tricolor
  red/green/blue laser diode based bidirectional signal remodulation
  visible-light communication system,'' \emph{Photonics Res.}, vol.~6, no.~5,
  pp. 422--426, Apr. 2018.

\end{thebibliography}
	%
	%
	
	%
	
	%
	%
	
	
	
	
	

	
\end{document}